\documentclass[a4paper,12pt]{article}
\pdfoutput=1
\usepackage{jheppub}
\usepackage{dcolumn}
\usepackage{soul}
\usepackage[english]{babel}
\usepackage[utf8]{inputenc}
\usepackage{dcolumn}
\usepackage{soul}

\usepackage{graphicx}
\usepackage{bm,amsmath,amssymb}
\usepackage[mathscr]{eucal}
\usepackage{makeidx}
\usepackage{subfig}
\usepackage{amsmath}
\usepackage{amssymb}
\usepackage{amsthm}
\usepackage{mathrsfs}
\usepackage{graphicx}
\usepackage{fancyhdr}
\usepackage{array}
\usepackage{simplewick}
\usepackage{latexsym}
\usepackage[all]{xy}
\usepackage{enumerate}
\usepackage{dsfont}
\usepackage{slashed}
\usepackage{float}
\usepackage[titletoc]{appendix}
\usepackage{ulem}

\usepackage{verbatim}

\usepackage{tikz,tikz-3dplot}
\tdplotsetmaincoords{80}{45}

\tikzset{surface/.style={draw=black, fill=white, fill opacity=.6}}

\usepackage{tikz}
\usetikzlibrary{decorations.pathmorphing,patterns}

\newcommand{\be}{\begin{equation}}
\newcommand{\ee}{\end{equation}}
\newcommand{\bea}{\begin{eqnarray}}
\newcommand{\eea}{\end{eqnarray}}

\newcommand{\ben}{\begin{eqnarray}}
\newcommand{\een}{\end{eqnarray}}
\newcommand{\sech}{\rm sech}

\title{de Sitter versus anti-de Sitter in Horndeski-like gravity}

\author[a]{Fabiano F. Santos}
\author[b]{Behnam Pourhassan}
\author[c]{Emmanuel N. Saridakis}

\affiliation[a]{Departamento de Física, Universidade Federal do Maranhão, Campus Universitario do Bacanga, São Luís (MA), 65080-805, Brazil.\\Instituto de Física, Universidade Federal do  Rio de Janeiro, 
Rio de Janeiro-RJ, 21941-972 -- Brazil.}
\emailAdd{fabiano.ffs23@gmail.com}

\affiliation[b]{School of Physics, Damghan University, Damghan, 3671641167, Iran.}
\emailAdd{b.pourhassan@du.ac.ir} 

\affiliation[c]{Institute for Astronomy,  Astrophysics, Space Applications and 
Remote Sensing (IAASARS), National Observatory of Athens, Athens, Greece.}
\emailAdd{msaridak@noa.gr} 

\abstract{We present general solutions of Horndeski-like gravity that can interpolate between the de Sitter and anti-de Sitter regimes. In particular, we develop the first-order formalism with two scalar fields, and considering a black hole ansatz with flat slicing we investigate three different cases, namely exponential, vacuum, and smooth superpotential solutions, with no Minkowski extrema. Furthermore, with these solutions we show that a Renormalization Group flow is established, and we obtain a turnaround in the warp factor, where the transition is bounded by the area low. We discuss the ideal regimes to trap gravity, which are constructed using the holographic function, which provides stable and unstable regimes to localize gravity. Finally, we show that no ghost appear and  that  the matter sector that violates the $c$-theorem is physical.
} 

\begin{document}
	\maketitle

\section{Introduction}

The Anti-de Sitter/Conformal Field Theory (AdS/CFT) correspondence \cite{Maldacena:1997re} is a powerful tool to investigate strongly coupled conformal field theory, where the macroscopic properties of strongly coupled matter are treated through the non-perturbative methods (AdS/CFT correspondence) 
\cite{Gubser:1998bc,Witten:1998qj,Hartnoll:2008kx,Hartnoll:2009sz}. The weakly coupled is dual to classical anti-de Sitter (AdS) gravity. This gauge/gravity duality determines the conformal anomaly of the CFT by means of the Fefferman-Graham (FG) expansion in bulk gravity 
\cite{Henningson:1998gx,Henningson:1998ey,Imbimbo:1999bj}, while the holographic conformal anomaly arises as a result of the boundary energy-momentum tensor trace not disappearing in vacuum \cite{Imbimbo:1999bj}.

On the asymptotic AdS spacetime side, the holographic correspondence provides a picture of the theory space structure and its connections via string theory/supergravity solutions that correspond to Quantum Field Theory (QFT) Renormalization Group (RG) flows \cite{Girardello:1998pd,Ghosh:2017big}. 
However, there is a concrete framework for understanding QFT mapping 
\cite{Douglas:2010rc,Furuuchi:2005qm}.

In the holographic framework, quantum field theory (QFT) implies that the states of the higher energy modes are integrated along the renormalization group (RG) passing from a higher energy state to a lower energy one. In this sense, the degrees of freedom decrease irreversibly. In a recent investigation  
 on Horndeski theories for a general coupling constant \cite{CANTATA:2021ktz}, it was shown that a holographic theorem $a$ is established for a critical point  \cite{Li:2018kqp}. Thus, there are charges known as $a$-charges that measure the massless degrees of freedom of the CFT at the fixed points of the RG, and these charges at the fixed ultraviolet (UV) are always greater than or equal to those of the infrared (IR) fixed point.

Recently, de Sitter (dS) space has attracted attention in the holographic scene \cite{Susskind:2021esx,Shaghoulian:2022fop,Addazi:2021xuf}. However, such a space is still considered ``exotic'' according to holographic scenario,  despite its cosmological significance at early and late 
cosmological times. The dS space together with the associated quantum physics presents several puzzles, one of which is the question of the size of the cosmological constant and the fact that it appears dynamically unstable to quantum corrections \cite{Mukhanov:1996ak,Tsamis:1996qq}.

The theoretical framework of  weakly coupled, weakly curved strings, seems to be in conflict  with the dS solutions \cite{Danielsson:2018ztv}. Many attempts to find such solutions rely on a general structure that is difficult to control quantitatively \cite{Kachru:2018aqn}, and with other difficulties associated with (holographic) control of anti-branes. However, efforts to find controllable dS solutions in string theory were presented by \cite{Cordova:2018dbb}, while new possible forms have been proposed based on the idea of the braneworld \cite{Ghosh:2018fbx}. In the following these solutions will be controlled by the parameters of Horndeski-like theory, as well as by the solutions arising from first-order formalism. 

Although holographic representations of AdS are well-defined, there is no completely concrete 
representation of de Sitter space \cite{Susskind:2021esx}. Thus, motivated by this contrast, Susskind presents specific principles and a well-defined example that realizes these principles, in the holographic framework for static corrections (SP) \cite{Susskind:2021omt,Susskind:2021dfc}. In particular, one assumes that there is a unitary Hamiltonian quantum mechanics of a static (dS) patch, where the degrees of freedom are located on the stretched horizon. In this way, the entanglement, chaos, and complexity roles 
are used to derive the necessary requirements, which are very different from those for AdS for a quantum system to be dual to the dS space. Although these requirements are met by a non-standard threshold, they are perfectly defined in the Sachdev-Ye-Kitaev (SYK) system \cite{Berkooz:2020uly}.

 In this work we will perform an investigation based on the work of \cite{Kiritsis:2013gia,Kiritsis:2019wyk}, considering Horndeski-like gravity coupled with two scalar fields and  using the first-order formalism 
\cite{Saridakis:2016ahq,Saridakis:2016mjd,Brito:2018pwe,Santos:2022lxj,
Santos:2019ljs,Santos:2021guj, Santos:2022uxm,Banerjee:2022ynv} to study the RG 
flow \cite{northe}. Furthermore, we will establish the dual gravitational description of the RG flow, obtaining an interpolation between de Sitter (dS) and AdS regimes.

In particular, in our context we will work with a black-hole ansatz with flat slicing in Horndeski-like gravity. Such ansatz does not have Minkowski extrema, nevertheless the solutions through the first-order formalism do have scalar superpotential interpolations between asymptotic dS and asymptotically 
AdS spacetimes, when the parameter $\gamma$ of Horndeski-like gravity changes from values $\gamma<1$ to $ \gamma>1$. In this sense, the dS space appears as a braneworld of the type discussed by Randall and Sundrum \cite{Randall:1999vf}, which is delimited by pode and antipode. These two regions for our dS-space 
gravity localization scenario  are responsible for locating and trapping gravity. Thus, with a maximum number of bit-threads squeezed without overlap between the asymmetric branches, the entanglement entropy can be calculated \cite{Shaghoulian:2022fop}. The surface on which it is computed as a minimum 
area defines the bottleneck that controls the maximum number of bit-threads.

 Additionally, we will discuss the minimum area to locate gravity both in dS space and in AdS space through the holographic entanglement \cite{Fabiano-JHAP}, which is interpolated due to the evolution of the $\gamma$ parameter. A feature in this interpolation of our holographic coordinate is UV dS branes that can respawn when viewed from an incompatible infrared brane \cite{Karch:2020iit} (in the gauge/gravity duality \cite{Maldacena:1997re,Gubser:1998bc,Witten:1998qj} or the domain wall/QFT 
correspondence \cite{Boonstra:1998mp,Bazeia:2006ef}, there is the possibility of considering the warp factor of a space-time geometry as a scale of the energy of a dual-field holographic theory at its boundaries). Moreover, we will evaluate the behavior of these branes with a minimum area trapping gravity through the $\beta$-function of the QFT boundary \cite{Kiritsis:2013gia,Kiritsis:2019wyk} in 
terms of the superpotential that has dependence on the  Horndeski-like coupling parameters, which provides greater freedom to find the solutions that lead to the GR flows.  

The paper is organized as follows. In Sec. \ref{V1} we present Horndeski-like gravity. In Sec. \ref{V2}  we develop the first-order formalism in five dimensions for black hole ansatz with flat slicing, and we extract the solutions that interpolate between dS and AdS space. In Sec. \ref{null} we present the analysis of the null energy condition through the null boundaries to the Horndeski-like gravity, and we compute the holographic entanglement entropy. In Sec. \ref{V3} we analyze the holographic scenario through the first-order formalism to show the ultraviolet (UV) fixed and the infra-ref (IR) fixed point using the $\beta(\phi)$ function, this function is a strong mechanism to check regions where the gravity can be localized and with this, we present some cases of gravity localization. Finally, in 
Sec. \ref{V4}, we summarize and conclude.

\section{The setup}\label{V1}

In this section we present a Horndeski-like gravity with two scalar fields \cite{Santos:2022uxm}, a scenario that accepts analytic solutions. We start with the action 
\begin{equation}
S[g_{\sigma\rho},\phi]=\int_{\mathcal{M}}{\sqrt{-g}d^{5}x(\mathcal{L}_{H}-V(\phi,\chi))}+S_{GH},\label{1}
\end{equation}
where
\begin{equation}
\mathcal{L}_{H}=\kappa(R-2\Lambda)-\frac{1}{2}(\alpha g_{\sigma\rho}-\gamma G_{\sigma\rho})\nabla^{\sigma}\phi\nabla^{\rho}\phi-\frac{1}{2}\nabla_{\mu}\chi\nabla^{\mu}\chi.\label{1.1}
\end{equation}
As we observe, we include a non-minimal coupling controlled by the $\gamma$ parameter (with dimensions $( 
mass)^{-2}$), and $\kappa=16\pi G_{N}$ with $G_{ N}$ the Newton gravitational constant. The scalar field has dimension $(mass)^{2}$ and the parameter $\alpha$ is dimensionless. Since $\phi$ and $\chi$ appear in the action only through a derivative, there is constant displacement symmetry associated with $\phi$ and $\chi$, implying that $\phi$ and $\chi$ are axionic. $S_{GH}$ is the Gibbons-Hawking term dependent on the 
parameter $\gamma$, namely
\begin{eqnarray}
&&S_{GH}=-2\kappa\int_{\partial\mathcal{M}}{d^{4}x\sqrt{\bar{\gamma}}\mathcal{L}
_{b}}+2\kappa\int{d^{4}x\sqrt{\bar{\gamma}}\mathcal{L}_{ct}},\label{2}
\end{eqnarray}
with
\begin{eqnarray}
&&\mathcal{L}_{b}=K^{({\bar{\gamma}})}-\Sigma^{(\bar{\gamma})}+\frac{\gamma}{4}\left(\nabla_{\mu}\phi\nabla_{\nu}\phi\, n^{\mu}n^{\nu}-(\nabla\phi)^{2}\right)K^{(\bar{\gamma})}+\frac{\gamma}{4}\nabla^{\mu}\phi\nabla^{\nu}\phi K^{(\bar{\gamma})}_{\mu\nu},\label{3}\\
&&{\cal L}_{ct}=c_{0}+c_{1}R+c_{2}R^{ij}R_{ij}+c_{3}R^{2}+b_{1}(\partial_{i}\phi\partial^{i}\phi)^{2}.\label{4}
\end{eqnarray}
Here, $\mathcal{L}_{b}$ corresponds to the Gibbons-Hawking $\gamma$-dependent term associated with Horndeski-like gravity, $n^{\mu}$ is an outward pointing unit normal vector to the boundary, 
$K_{\mu\nu}=\bar{\gamma}^{\beta}_{\mu}\nabla_{\beta}n_{\nu}$ is the extrinsic curvature, $K^{(\bar{\gamma})}=\bar{\gamma}^{\mu\nu}K^{({\bar{\gamma}})}_{\mu\nu}$ is the trace of the extrinsic curvature, and $\bar{\gamma}_{\mu\nu}$ is the induced metric on the boundary $r\to\infty$. Finally, the Lagrangian ${\cal L}_{ct}$ is related to the boundary counterterms, and since they do not affect the bulk 
dynamics they will be neglected.

In the dual QFT, the $d$-dimensional Minkowski space-time is defined, which is the limit of the $(d+1)$-spacetime for which Einstein's scalar theory is defined. In this gravitational frame of reference, we have that the saddle point of the ground state of the QFT is related via holography to the invariant 
solutions of Poincaré and Einstein's scalar theory. To  obtain these solutions one can always work in the so-called domain wall coordinate system \cite{Brito:2018pwe,Santos:2022lxj,Santos:2019ljs,Santos:2022uxm,Santos:2021guj}
:
\begin{equation}
ds^{2}=g_{\sigma\rho}dx^{\sigma}dx^{\rho}=e^{2A(u)}g_{\mu\nu}dx^{\mu}
dx^{\nu}-du^{2}\label{metric} ,
\end{equation}
where Latin indices $\sigma,\rho$$\,\in[ $0$, $1$, $2$, $3$, $4$ ]$ run on the bulk and Greek indices $\mu,\nu$$\,\in[ $0$, $1$, $2$, $3$ ]$ run along the braneworld coordinates (i.e. $u$ is the 
holographic coordinate). The (\ref{metric}) is manifested ISO(d) invariant, where the dynamic variable is the scale factor $e^{A}$ of the Minkowski slices. We use dots to indicate  derivatives with respect to the holographic coordinate, while the derivatives with respect to the scalar field are indicated with a prime.

{\underline{Gravity side}}: From the gravitational part we have:

\begin{eqnarray}
E_{\sigma\rho}&=&-\frac{2}{\sqrt{-g}}\frac{\delta S^{\mathcal{M}}}{\delta g^{\sigma\rho}}\,,\nonumber\\ 
E_{\phi}&=&-\frac{2}{\sqrt{-g}}\frac{\delta S^{\mathcal{M}}}{\delta\phi}\,,\nonumber\\ E_{\chi}&=&-\frac{2}{\sqrt{-g}}\frac{\delta S^{\mathcal{M}}}{\delta\chi} \,,\nonumber\\ F_{\phi}&=&-\frac{2}{\sqrt{-\bar{\gamma}}}\frac{\delta S^{\partial\mathcal{M}}}{\delta\phi} \,,
\end{eqnarray}
where
\begin{equation}
E_{\sigma\rho}=G_{\sigma\rho}+\Lambda g_{\sigma\rho}-\frac{1}{2k}T_{\sigma\rho}=0,\label{S1}
\end{equation}
with $T_{\sigma\rho}=\alpha T^{(1)}_{\sigma\rho}-g_{\sigma\rho}V(\phi)+\gamma 
T^{(2)}_{\sigma\rho}$, and with
\begin{eqnarray}
&&
\!\!\!\!\!\!\!\!\!\!\!\!\!\!\!\!\!
T^{(1)}_{\sigma\rho} 
=\nabla_{\sigma}\phi\nabla_{\rho}\phi-\frac{1}{2}g_{\sigma\rho}\nabla_{\lambda}
\phi\nabla^{\lambda}\phi+\nabla_{\sigma}\chi\nabla_{\rho}\chi-\frac{1}{2}g_{\sigma\rho}\nabla_{\lambda}
\chi\nabla^{\lambda}\chi,\\
&&\!\!\!\! \!
\!\!\!\!\!\!\!\!\!\!\!\!T^{(2)}_{\sigma\rho}=\frac{1}{2}\nabla_{\sigma}
\phi\nabla_{\rho} \phi R-2\nabla_{\lambda}\phi\nabla_{(\sigma}\phi 
R^{\lambda}_{\rho)}-\nabla^{\lambda}\phi\nabla^{\tau}\phi 
R_{\sigma\lambda\rho\tau}\nonumber\\ 
&&-(\nabla_{\sigma}\nabla^{\lambda}\phi)(\nabla_{\rho}\nabla_{\lambda} 
\phi)+(\nabla_{\sigma}\nabla_{\rho}\phi) 
\square\phi+\frac{1}{2}G_{\sigma\rho}(\nabla\phi)^{2}\nonumber\\
							&&- 
g_{\sigma\rho}\left[-\frac{1}{2}(\nabla^{\lambda}\nabla^{\tau}\phi)(\nabla_{
\lambda}\nabla_{\tau}\phi)+\frac{1}{2}(\square\phi)^{2}-(\nabla_{\lambda}
\phi\nabla_{\tau}\phi)R^{\lambda\tau}\right]. \label{S2}
\end{eqnarray}
Furthermore, we have 
\begin{eqnarray}
E_{\phi}&=&\nabla_{\mu}\left[\left(\alpha g^{\mu\nu}-\gamma G^{\mu\nu}\right)\nabla_{\nu}\phi\right]-\frac{dV(\phi,\chi)}{d\phi}\,,\label{L11}\\
E_{\chi}&=&\ddot{\chi}(u)+4\dot{A}(u)\dot{\chi}(u)-\frac{dV(\phi,\chi)}{d\chi}=0\\
F_{\phi}&=&-\frac{\gamma}{4}(\nabla_{\mu}\nabla_{\nu}\phi n^{\mu}n^{\nu}-(\nabla^{2}\phi))K-\frac{\gamma}{4}(\nabla_{\mu}\nabla_{\nu}\phi)K^{\mu\nu}\,,\label{L12}
\end{eqnarray}
and note that $E_{\phi}=F_{\phi}$ due to the Euler-Lagrange equation 
\cite{Santos:2021orr,Sokoliuk:2022llp,Santos:2023flb}.

For null $\phi$, $\chi$ and $V(\phi)=const.$ as described by \cite{Li:2018kqp}, the equations for $E_{\sigma\rho}$, $E_{\phi}$ and $F_{\phi}$ admit an AdS vacuum of maximum symmetry  with $G_{\sigma\rho}=-\Lambda_{0}g_{\sigma\rho}$, and for $\mathcal{L}_{H}$ in the (\ref{1}) the absence of phantom excitation requires that $\alpha+\gamma\Lambda_{0}\geq\,0$, with equality corresponding to the critical point. At a critical point of the coupling $\gamma G_{\sigma\rho}\nabla^{\sigma}\phi\nabla^{\rho}\phi$, there is an almost AdS spacetime for which $\phi$ is not zero, and the integration constant contributes to the effective cosmological constant.

{\underline{Boundary side}}:
On the boundary side $\partial\mathcal{M}$, following \cite{Santos:2021orr,Sokoliuk:2022llp,Santos:2023flb} we have:
\begin{eqnarray}
K_{\mu\nu}-\bar{\gamma}_{\mu\nu}(K^{(\bar{\gamma})}-\Sigma^{(\bar{\gamma})})-\frac{\gamma}{4}H_{\mu\nu}=\kappa {\cal S}^{\partial\mathcal{M}}_{\mu\nu}\,,\label{L7}
\end{eqnarray}
where
\begin{eqnarray}
&&H_{\mu\nu}\equiv(\nabla_{\mu}\phi\nabla_{\nu}\phi\, n^{\mu}n^{\nu}-(\nabla \phi)^2) (K_{\mu\nu}-\bar{\gamma}_{\mu\nu}K)-(\nabla_{\mu}\phi\nabla_{\nu}\phi)K\,,\label{L8}\\
&&{\cal S}^{\partial\mathcal{M}}_{\alpha\beta}=-\frac{2}{\sqrt{-\bar{\gamma}}}\frac{\delta S^{\partial\mathcal{M}}_{mat}}{\delta \bar{\gamma}^{\alpha\beta}}\,.\label{L9} 
\end{eqnarray}
Considering the matter stress-energy tensor on $\partial\mathcal{M}$ as a constant (i.e. ${\cal S}^{\partial\mathcal{M}}_{\alpha\beta}=0$), we can write
\begin{eqnarray}
K_{\mu\nu}-\bar{\gamma}_{\mu\nu}(K^{(\bar{\gamma})}-\Sigma^{(\bar{\gamma})})-\frac{\gamma}{4}H_{\mu\nu}=0\,.\label{L10}
\end{eqnarray}

\section{Black hole ansatz with flat slicing}\label{V2}

The idea in this section is to study solutions that interpolate between asymptotic dS and asymptotic AdS space-times in Horndeski-like gravity. Note that solutions of this type were extracted numerically in  
\cite{Kiritsis:2019wyk}. In our case, we develop the first-order formalism with two axionic fields to obtain analytical solutions, focusing on solutions that interpolate between dS and AdS space-time. We consider the coordinate system that allows for the coordinate $u$ to change from space-like (asymptotically AdS spacetimes) to time-like (asymptotically dS spacetimes), and the dynamical variable will be the blackness function $f(u)$. When $f(u)$ vanishes yields a horizon, on either side of which $f$ has a different sign. In summary, a solution that passes through a horizon, for $f$ in the ansatz, exchanges $u$ from space-like to time-like and vice-versa. Hence, we consider the ansatz
\begin{eqnarray}
ds^{2}=\frac{du^{2}}{f(u)}+e^{2A(u)}[-f(u)dt^{2}+dx^{2}+dy^{2}+dz^{2}].\label{metricflat}
\end{eqnarray}

We proceed to the equations of motion for metric \eqref{metricflat}, combined with the first-order formalism:
\begin{eqnarray}\label{eq:1}
&&\dot{A}(u)=-\frac{1}{3}W(\phi,\chi),\label{eq:1}\\
&&\dot{\phi}(u)=cW_{\phi},\label{eq:1.1}\\
&&\dot{\chi}(u)=cW_{\chi}.\label{eq:1.2}
\end{eqnarray}
 In that case, considering the Horndeski-like gravitational sector with $\psi(u)=\dot{\phi}(u)$, combining the $tt$-component with $xx$, $yy$, or $zz$-components, we have
\begin{eqnarray}\label{eq:2}
0&=&8\kappa\Lambda+4V(\phi,\chi)+12\kappa\dot{A}\dot{f}+12\gamma\,f^{2}\dot{A}^{2}\psi^{2}+12\gamma\,f^{2}\dot{A}\psi\dot{\psi}+6\gamma\,f^{2}\ddot{A}\psi^{2}\nonumber\\
&+&2\alpha\,f\psi^{2}+9\gamma\,f\dot{f}\dot{A}\psi^{2}+48\kappa\,f\dot{A}^{2}+24\kappa\,f\ddot{A}+2f\dot{\chi}^{2},
\end{eqnarray}
while for the $rr$-components we have
\begin{eqnarray}\label{eq:3}
V(\phi,\chi)=&-&9\gamma\,f^{2}\dot{A}^{2}\psi^{2}-3\kappa\dot{A}\dot{f}-12\kappa\,f\dot{A}^{2}\nonumber\\
&+&\frac{1}{2}\alpha\,f\psi^{2}-\frac{9}{4}\gamma\,f\dot{f}\dot{A}\psi^{2}+\frac{1}{2}f\dot{\chi}^{2}-2\kappa\Lambda.
\end{eqnarray}
On the other hand, the equation describing the scalar field dynamics is given as
\begin{eqnarray}\label{eq:sca}
&-&2\alpha\,f\dot{\psi}+12\gamma\,f^{2}\dot{A}^{2}\dot{\psi}+3\gamma\,f\dot{f}\dot{A}\dot{\psi}+\dot{f}\psi(-2\alpha+3\gamma\dot{f}\dot{A})+24\gamma\,f^{2}\dot{A}\psi(2\dot{A}^{2}+\ddot{A})\nonumber\\
&+&\dot{f}\psi[36\gamma\dot{f}\dot{A}^{2}+3\gamma\dot{f}\ddot{A}+\dot{A}(-8\alpha+3\ddot{f})]+2\frac{dV(\phi,\chi)}{d\phi}=0.
\end{eqnarray}
Firstly, combining the $tt$-component with the $rr$-component, and using (\ref{eq:1}) we find (with $c=\frac{1}{2}$):
\begin{equation}
\gamma\,WW_{\phi\phi}+\frac{\gamma\,W^{2}_{\phi}}{2}+\frac{4\gamma}{3}W^{2}-\frac{(2\alpha+8\kappa)}{f}-2\frac{W^{2}_{\chi}}{fW^{2}_{\phi}}=0,\label{eq:w0}
\end{equation}
where $W_{\phi}=dW/d\phi$ and $W_{\chi}=dW/d\chi$. Note that in (\ref{eq:w0}) the limit $\gamma\to0$  gives $W_{\chi}=\sqrt{-(\alpha+4\kappa)}W_{\phi}$, and this equation can be satisfied by $W(\phi,\chi)=e^{a\phi+b\chi}$ with $b=a\sqrt{-(\alpha+4\kappa)}$, and with $a$ and $b$ constants. In this case equations (\ref{eq:1})-(\ref{eq:1.2}) lead to
\begin{eqnarray}
&&\phi(u)=a\ln(u),\label{phi}\\
&&\chi(u)=b\ln(u),\label{chi}\\
&&A(u)=-\frac{1}{3}\frac{u^{a^2+b^2+1}}{a^2+b^2+1}.\label{A}
\end{eqnarray}
 
In usual Einstein gravity one can consider superpotential examples in order to find symmetric bent brane solutions \cite{Bazeia:2006ef}. Such configurations are in four-dimensions with AdS geometry, which are 
holographically dual to the field theory and exhibit a weakly coupled regime at high energy. In order to construct similar solutions in our model, namely solutions that provide a flow starting in the
dS maximum and ending in an AdS minimum, we need to consider $\gamma\neq\,0$.

We start  by re-writing 
(\ref{eq:w0}) as
\begin{equation}
WW_{\phi\phi}+\frac{W^{2}_{\phi}}{2}+\frac{4}{3}W^{2}-\frac{\sigma}{f}-2\frac{W^{2}_{\chi}}{\gamma\,fW^{2}_{\phi}}=0,\label{eq:w1}
\end{equation}
and combining the $rr$ equations with (\ref{eq:sca}) and (\ref{eq:1}) we find
\begin{equation}
WW_{\phi\phi}+\frac{W^{2}_{\phi}}{2}+\frac{4}{3}W^{2}-\frac{\sigma}{f}-\frac{3}{4\gamma\,f}\frac{W_{\chi}W_{\chi\phi}}{WW_{\phi}}=0,\label{eq:w2}
\end{equation}
where $\sigma=(2\alpha-8\kappa)/\gamma$. Hence, comparing (\ref{eq:w1}) and (\ref{eq:w2}) we deduce the following constraint on the superpotential:
\begin{equation}
\frac{3}{8}\frac{W_{\chi\phi}}{W}=\frac{W_{\chi}}{W_{\phi}}.\label{15}
\end{equation}
With this technique, we explore the classification of de Sitter (dS) and Anti-de Sitter (AdS) where the dS regime is part of the scalar field space for $V(\phi,\chi)>0$ and AdS regime is where $V(\phi,\chi)<0$.

\subsection{Case A: Exponential superpotential}\label{case1}

\begin{figure}[!ht]
\begin{center}
\includegraphics[scale=0.6]{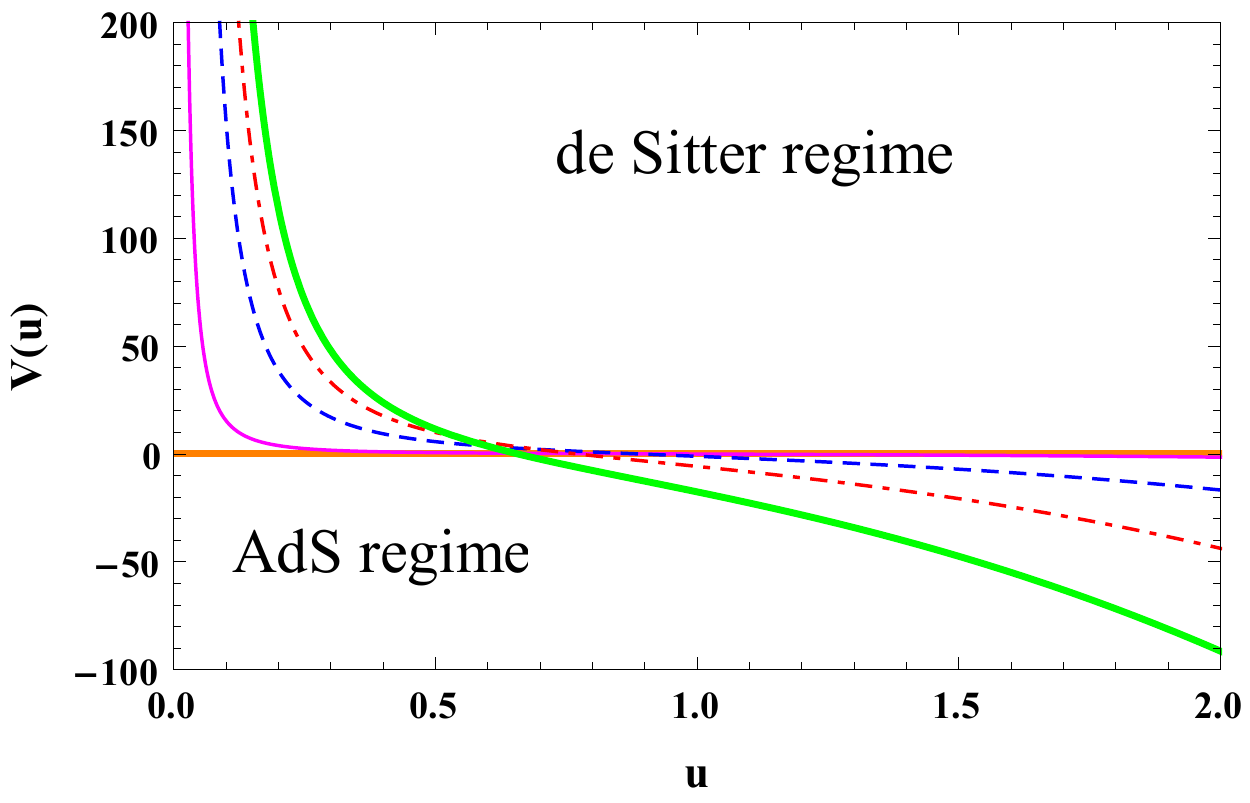}
\caption{{\it{The scalar potential (\ref{eq:3}) in the case of exponential 
superpotential 
$W(\phi,\chi)=e^{\sqrt{\frac{8}{3}}(\phi+\chi)}$, for 
 $\Lambda=0$, $\kappa=1/4$, $\alpha=8/3$, $\gamma=0.1$ (solid - pink), 
$\gamma=1$ (dashed - blue), $\gamma=2$ (dot-dashed - red) and $\gamma=3$ 
(thick - green), respectively.}}}\label{p}
\label{planohwkhz}
\end{center}
\end{figure}

We first consider the  simplest superpotential $W(\phi,\chi)=e^{\sqrt{\frac{8}{3}}(\phi+\chi)}$. In this case, solutions satisfying (\ref{eq:1}), alongside  constraint (\ref{15}), are given by  
\begin{eqnarray}
&&\phi(u)=-\frac{3}{4}\ln(u),\label{16}\\
&&\chi(u)=-\frac{3}{4}\ln(u),\label{17}\\
&&A(u)=\frac{1}{4}\ln(u).\label{18}
\end{eqnarray}
Inserting (\ref{16})-(\ref{18}) into (\ref{eq:w1}) we acquire the form of $f(u)$ as
\begin{eqnarray}
f(u)=\dfrac{3e^{3\sqrt{\frac{8}{3}}\ln(u)}}{16\left(\sigma-\frac{2}{\gamma}
\right)}.\label{19} 
\end{eqnarray}
Thus, (\ref{eq:3}) can now provide the behavior of the scalar potential, which is depicted in Fig. \ref{p}. As we can see, the direction of the flows is the direction in which the scalar potential in 
terms of $u$ decreases. In particular, the flow  starts in the dS regime ($V(u)>0$) and eventually results to an extremum of the AdS regime ($V(u)<0$). This transition from the dS to AdS regime is induced by the 
$\gamma$ parameter.   

\subsection{Case B: Vacuum solution}\label{case2}

\begin{figure}[!ht]
\begin{center}
\includegraphics[scale=0.6]{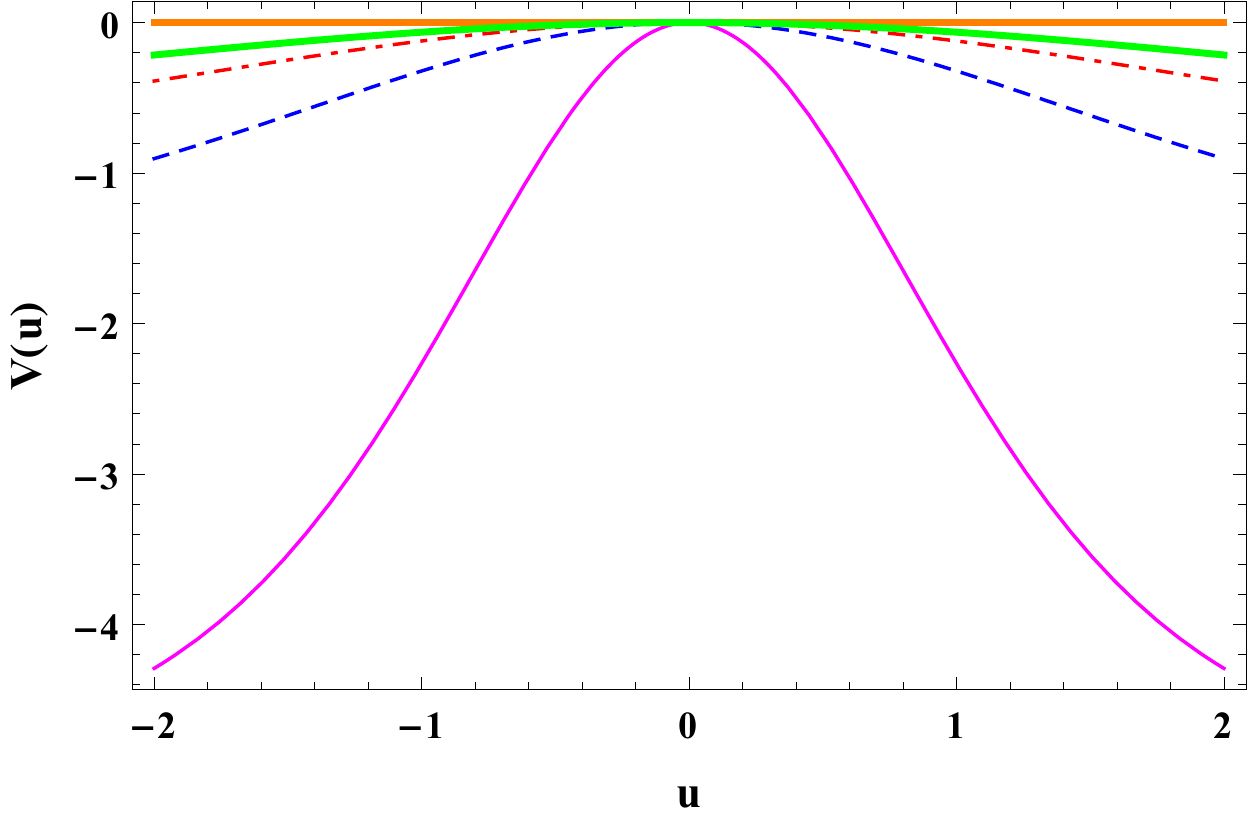}
\caption{{\it{The scalar potential (\ref{20}) for the vacuum solution with 
$W_{0}=\sqrt{3(\sigma+2/\gamma)}/4$, for  $\Lambda=0$, 
$\kappa=1/4$, $\alpha=8/3$, $\gamma=0.1$ (solid - pink), $\gamma=1$ (dashed -
blue), $\gamma=2$ (dot-dashed - red) and $\gamma=3$ (thick -  green).}}}
\label{p1}
\label{planohwkhz}
\end{center}
\end{figure}

A different   configuration that satisfies     (\ref{eq:w1}) and 
(\ref{eq:w2}) is $W_{0}=\sqrt{3(\sigma+2/\gamma)}/4$. For this case, we have 
$\phi,\chi=const.$ where the  warp factor can be found as 
$A=-(1/3)W_0\,u$, and thus this solution is the AdS$_5$ vacuum solution.  Points 
at which 
$W_{\phi}(\phi,\chi)=W_{\chi}(\phi,\chi)=0$ are known as critical points 
because 
these are the only points where the solutions of the superpotential equation 
can 
have singularities. Indeed, the behavior of $W(\phi,\chi)=0$  at a critical 
point is dictated by whether $V_{\phi}(\phi,\chi)=V_{\chi}(\phi,\chi)=0$ 
vanishes. The above solution  admits a critical point 
$\alpha+\gamma\Lambda_{0}\geq\,0$, where at this 
point the equality reduces Horndeski-like theory to Einstein gravity. The 
scalar potential (\ref{eq:3}) reduces to the form
\begin{eqnarray}
V(u)=-\frac{3}{4}\dot{f}\dot{A}-3f\dot{A}^{2};\quad f=\frac{3\sigma}{4W^{2}_{0}}.\label{20}
\end{eqnarray}
From  (\ref{20}) we can see that $\dot{A}(u)$ cannot increase (from the 
holographic RG flow side this is related to the holographic 
$c$-theorem \cite{Girardello:1998pd,Freedman:1999gp}). Finally, the above 
behavior is depicted in Fig. \ref{p1}.

\subsection{Case C: Smooth solution}\label{case3}

\begin{figure}[!ht]
\begin{center}
\includegraphics[scale=0.5]{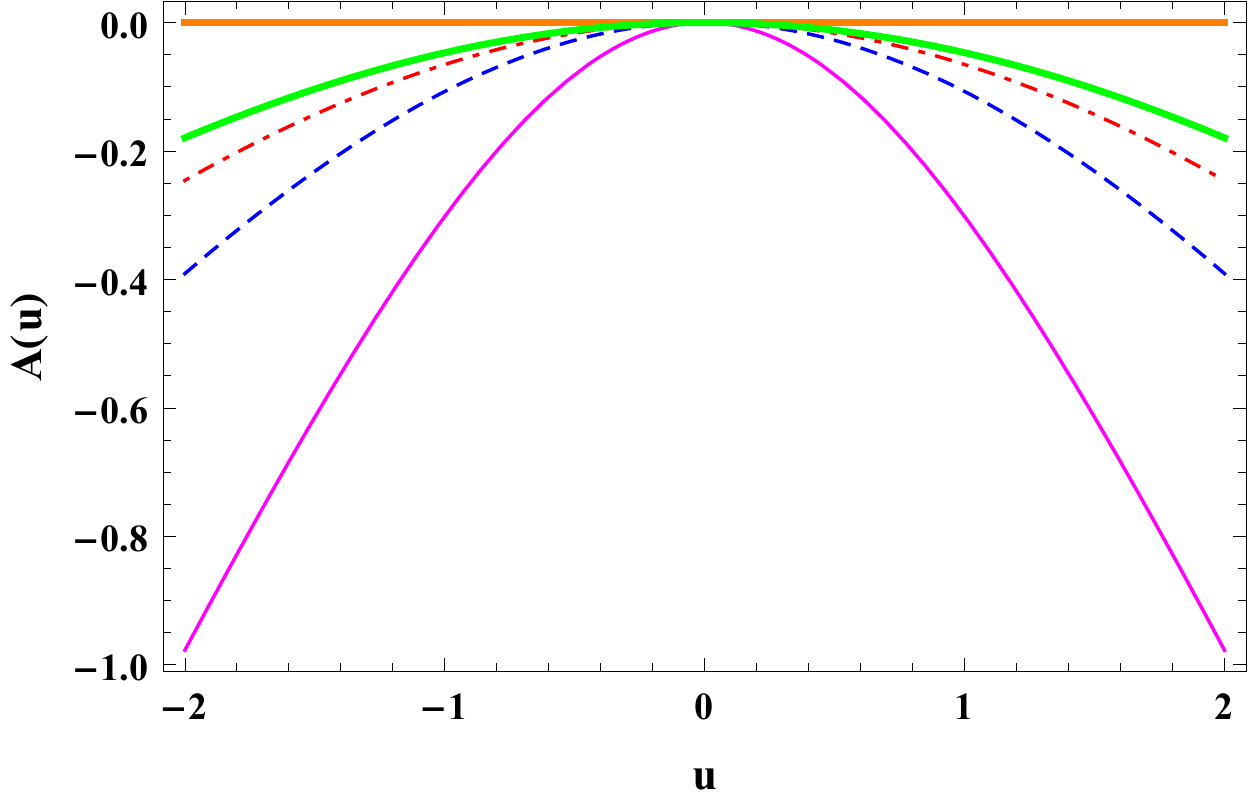}
\includegraphics[scale=0.5]{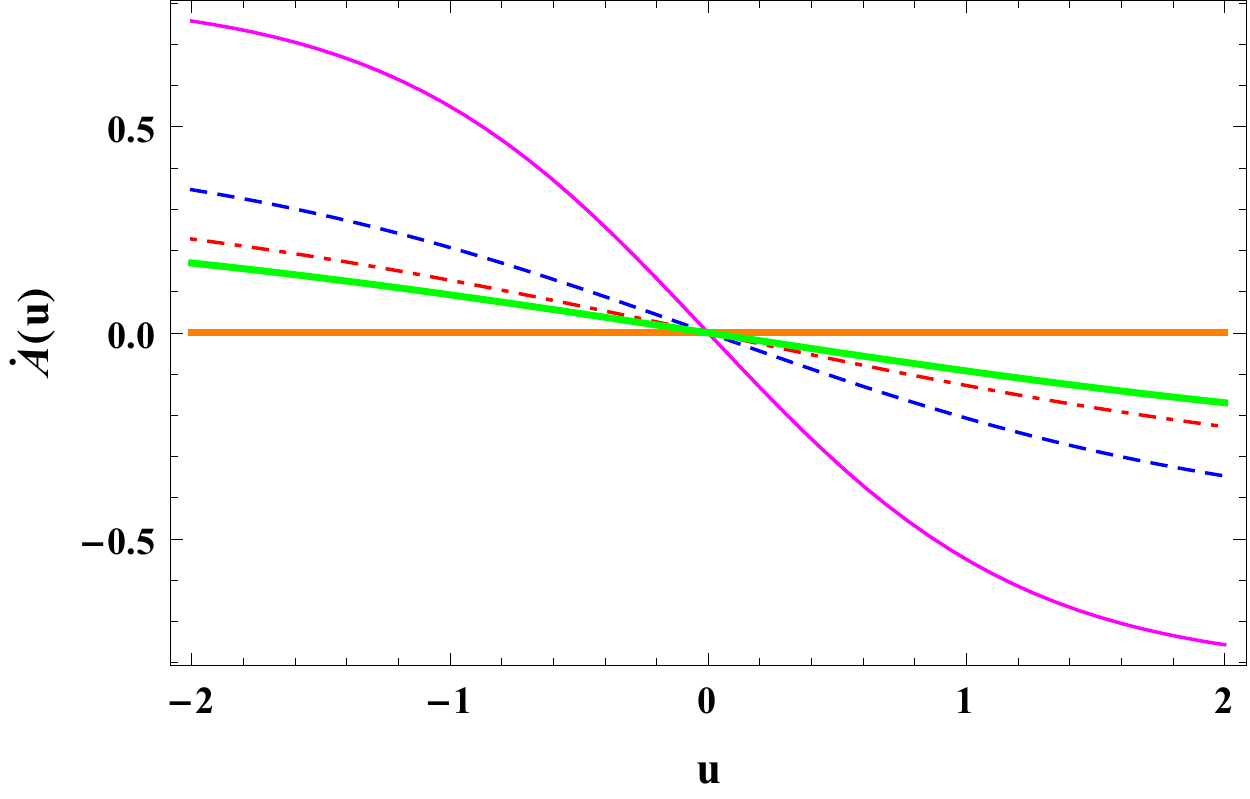}
\includegraphics[scale=0.5]{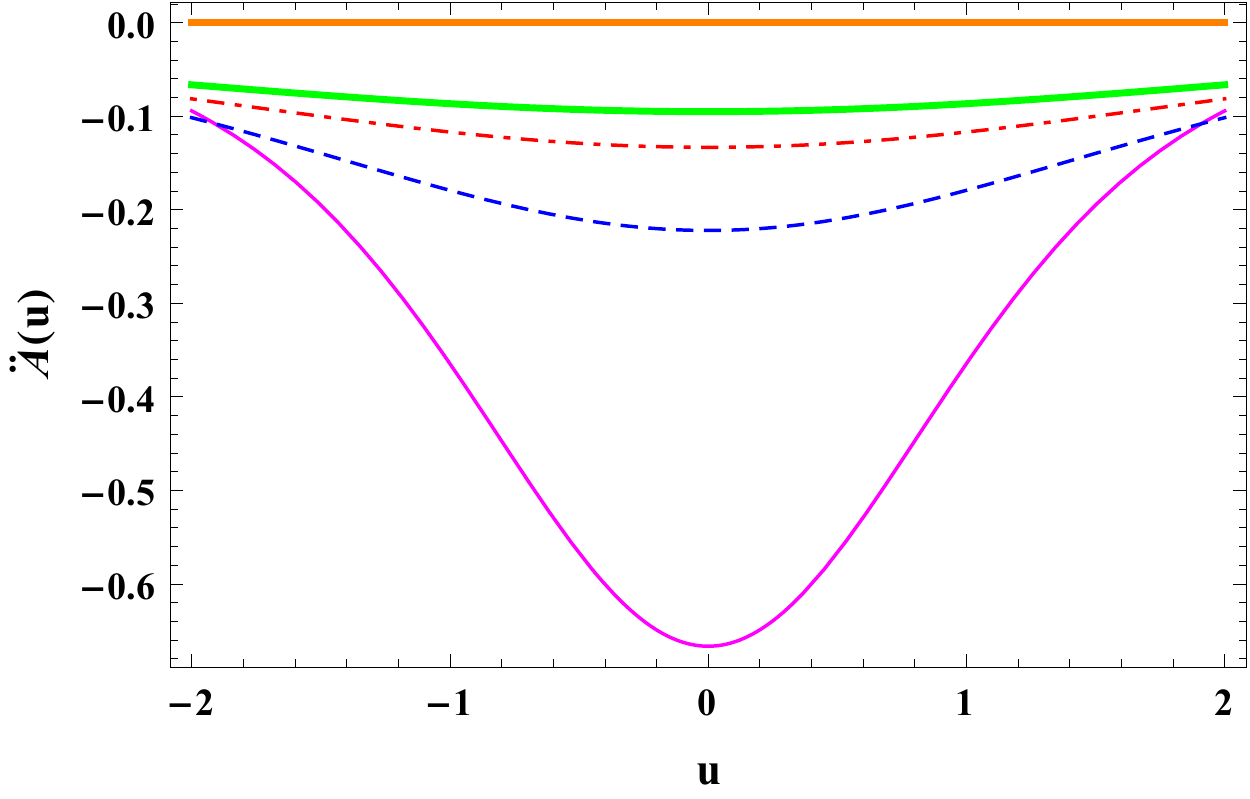}
\includegraphics[scale=0.5]{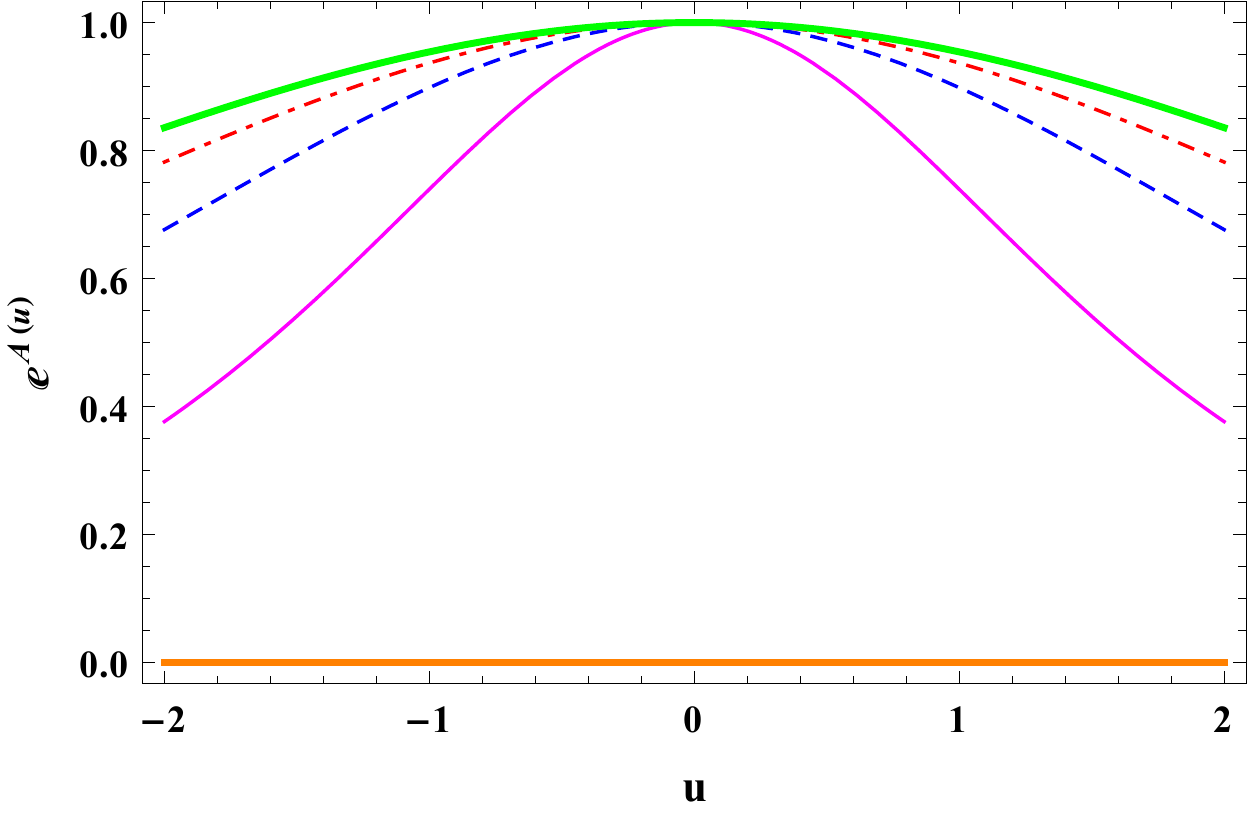}
\caption{{\it{The behavior of $A(u)=-\ln{\cosh{ku}}$, 
$\dot{A}(u)$, $\ddot{A}(u)$, and $e^{A(u)}$, for $\Lambda=0$, 
$\kappa=1/4$, $\alpha=8/3$, $\gamma=0.1$ (solid - pink), $\gamma=1$ (dashed - 
blue), $\gamma=2$ (dot-dashed - red) and $\gamma=3$ (thick - 
green).}}}
\label{p2}
\end{center}
\end{figure}

There are solutions  that are well known to localize gravity on the brane, with 
source given by a delta-function, patching together the AdS branches 
$A=\pm ku$ along with the  brane according to the solution $A=-k|u|$  
\cite{Randall:1999vf,csaki}. In our case, these assumptions are precisely the 
Randall-Sundrum scenario \cite{Randall:1999vf}, where 
$k=\sqrt{W_{0}}$ is related to the brane tension in the thin-wall limit. In 
order to address the issue of the energy distribution on the brane in a more 
transparent way, we smooth out this brane solution as $k|u|\to 
\ln{\cosh{ku}}$ \cite{csaki}, acquiring the smooth solution
\begin{equation}
A(u)=-\ln{\cosh{ku}}\approx-\frac{k^2}{2}u^{2},\label{21}
\end{equation} 
where $k\ll1$ as $\gamma\gg1$ (the smooth limit). In Fig. \ref{p2} we present 
$\dot{A}(u)$ and its derivatives, as well as the scale factor $e^{A(u)}$. As we 
can see, the flow originates from a dS maximum at $u=+2$ and ends at an AdS 
minimum at $u=-2$, while we verify that $\dot{A}(u)$ is decreasing. 

\begin{figure}[!ht]
\begin{center}
\includegraphics[scale=0.5]{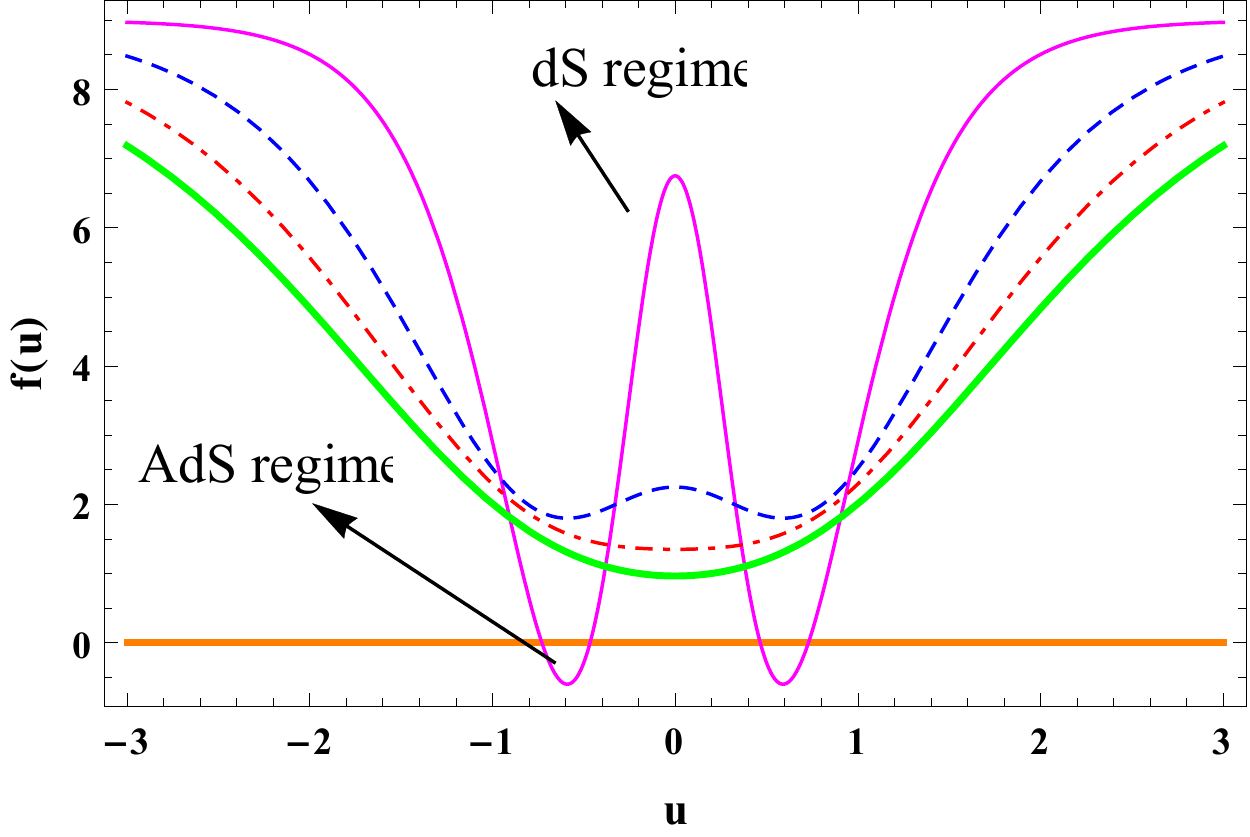}
\caption{{\it{The  profile of $f(u)$ for   
$\Lambda=0$, $\kappa=1/4$, $\alpha=8/3$, $\gamma=0.1$ (solid - pink), 
$\gamma=1$ (dashed - blue), $\gamma=2$ (dot-dashed - red) and 
$\gamma=3$ (thick - green).}
}}
\label{p3}
\end{center}
\end{figure}

\begin{figure}[!ht]
\begin{center}
\includegraphics[scale=0.5]{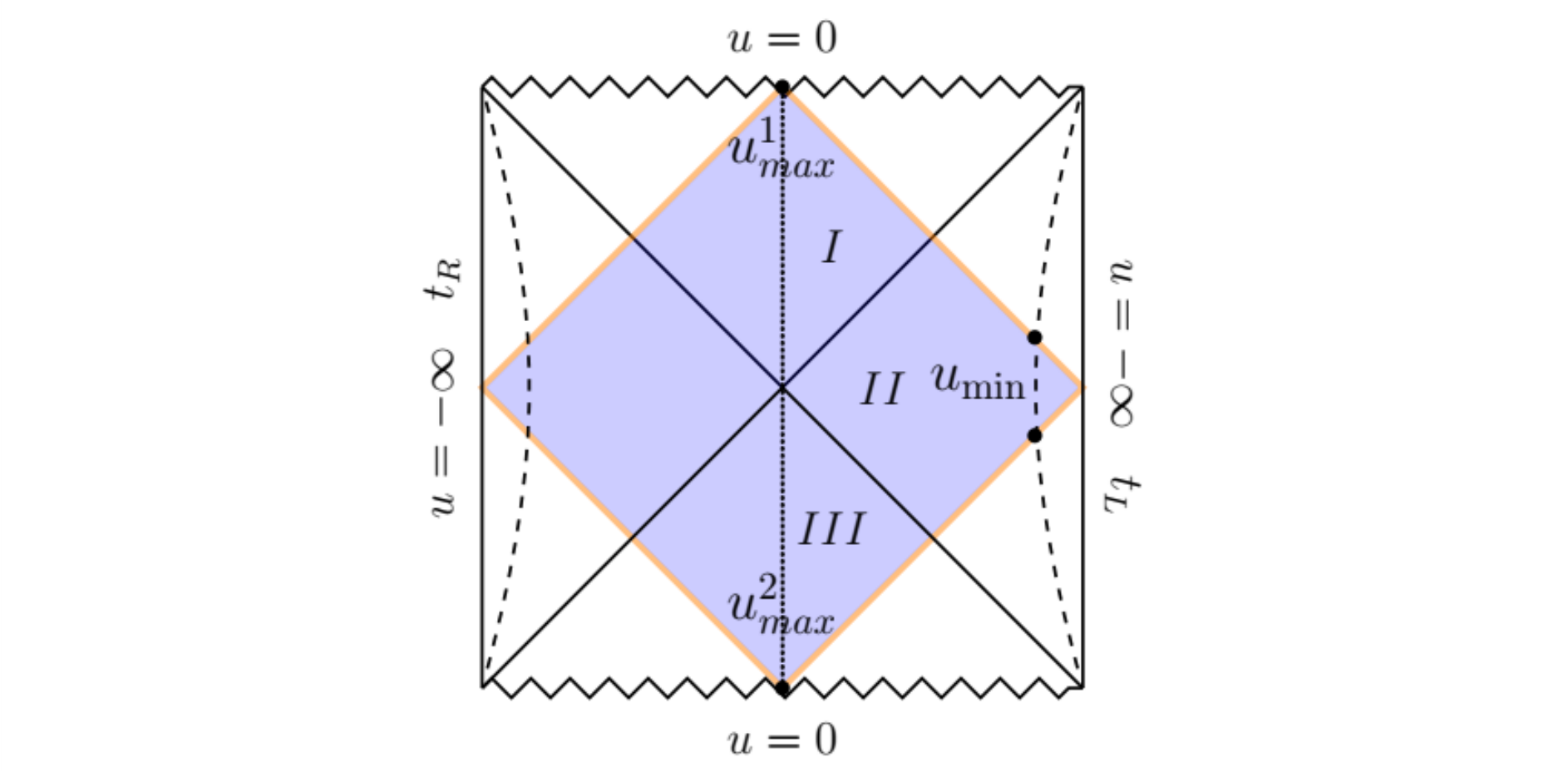}
\caption{{\it{The bulk conformal diagram corresponding to Fig. \ref{p3}, where  
 $u\to 0$ is the singular surface and $u\to\,-\infty$ is the asymptotic 
boundary surface. The black dashed curves correspond to UV cutoff surfaces at 
$u=u_{min}$, while $u^{1}_{max}$, $u^{2}_{max}$ are meeting points of null 
boundaries in the bulk.}}}\label{Null1}
\label{planohwkhz}
\end{center}
\end{figure}

 Finally, inserting $A(u)=-\ln{\cosh{ku}}$ into (\ref{eq:w1}) and (\ref{eq:w2}), we can extract the form of $f(u)$ that satisfy these equations with $W(u)=3k\tanh(ku)$ as: 
\begin{equation}
f(u)=\frac{3 \left(15 k^4 \text{sech}^4(k u)-12 k^4  \text{sech}^2(k u)-8 k^2 
\text{sech}^2(k u)+8 k^2\right)}{2 
\left(\sigma-\frac{2}{\gamma}\right)},\label{22}
\end{equation} 
and in Fig. \ref{p3} we draw its profile. Hence, $f(u)$ exhibits a horizon on 
which it vanishes, and it has different signs around it ($u$ changes from 
space-like to time-like and vice-versa). Therefore, our solutions in the AdS 
regime will have an AdS boundary at $u\to\,-\infty$. Thus, we have a flow 
originating from a dS maximum at $u^{2}_{max}$, $u^{1}_{max}$, and ending in an 
AdS minimum at $u_{\mathrm{min}}$. This structure is depicted in Fig. 
\ref{Null1}.

In our setup, as we show through the technique of first-order formalism, the potentially regular solutions exist for these generic three solutions starting in a dS maximum and ending at an AdS minimum. In fact, this interpolation between dS and AdS is provided also using the null-energy condition. To deduce this last condition we need to work with the null boundaries conditions, which will be present in the following section.  

\section{Null boundaries in Horndeski-like gravity}\label{null}

\begin{figure}[!ht]
\begin{center}
\includegraphics[scale=0.5]{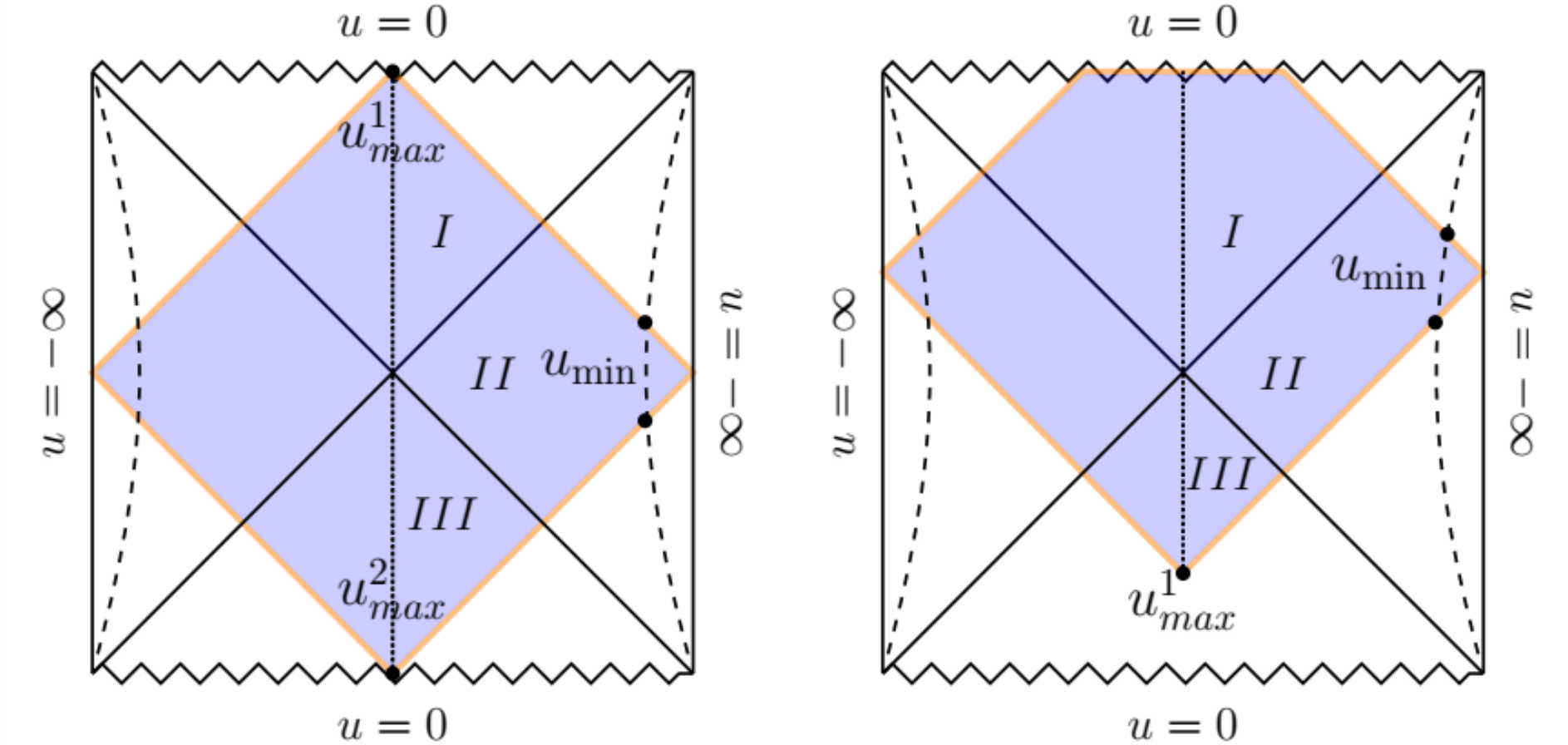}
\caption{{\it{Bulk conformal diagram at early ($t_R = t_L = \tau/2 = 0$) and late ($t_R = t_L = \tau/2 > 0$) times with the present singularity at the origin. The black dashed curves correspond to UV cutoff surfaces at $u=u_{min}$, while $u^{1}_{max}$ and $u^{2}_{max}$ are the intersecting points of null boundaries in the bulk.}
}}
\label{fig:1.0}
\label{planohwkhz}
\end{center}
\end{figure}

This section is devoted to presenting a complete discussion of the boundary term in the action functional of Horndeski-like gravity when the boundary includes null segments. We consider the affine parametrization for the null normals, where the null surface term vanishes. The Gibbons-Hawking term arises from the surface at the UV cutoff where we have a minimum. The total action is
\begin{eqnarray}
&&I_{total}=\int{\sqrt{-g}d^{5}x\mathcal{L}_{H}}-2\kappa\int{d^{4}x\sqrt{\bar{\gamma}}\mathcal{L}_{b}}+2\kappa\int{d^{4}x\sqrt{\bar{\gamma}}\mathcal{L}_{ct}}.\label{T1}
\end{eqnarray}
Since we only have null boundaries, it is more convenient to perform the calculation using the ingoing and outgoing coordinates like:
\begin{eqnarray}
v=t+u^{*}(u);\quad s=t-u^{*}(u),\label{Nu1}
\end{eqnarray}
where $u^{*}(u)=\int{e^{-A(u)}du}$ is a tortoise coordinate, with asymptotic behavior of the form
$\lim_{u\to\,-\infty}u^{*}(u)=u^{*}_{-\infty}$. The path includes  two UV cutoff surfaces near the asymptotic boundary regions at $u=u_{min}$, denoted by the black dashed curves in Fig. \ref{fig:1.0} 
($t_{L}$ and $t_{R}$ are the symmetric cutoffs \cite{Santos:2022lxj}). The inclusion of the two UV cutoff surfaces near the asymptotic boundary regions at $u=u_{min}$ are used to omit IR divergencies. Moreover, there are two intersecting points in the bulk due to the intersection with the future boundary hypersurface at $u=u^{1}_{max}$ and with the past one at $u=u^{2}_{max}$.  

It is important to mention that the null boundary is encoded in the time dependence of these points, which satisfies  
\begin{eqnarray}
\frac{t}{2}+u^{*}_{-\infty}-u^{*}(u^{1}_{max})=0,\quad 
\frac{t}{2}-u^{*}_{-\infty}+u^{*}(u^{2}_{max})=0,\label{Nu3} 
\end{eqnarray}
while  time evolution  is given by 
\begin{eqnarray}
\frac{du^{1}_{max}}{dt}=\frac{A(u^{1}_{max})}{2},\quad \frac{du^{2}_{max}}{dt}=-\frac{A(u^{2}_{max})}{2}.\label{Nu4}
\end{eqnarray}
In our prescription, the null boundaries of the right sector correspond to
\begin{eqnarray}
B_{1}:\frac{t}{2}=u^{*}(u)-u^{*}_{-\infty},\quad B_{2}:-\frac{t}{2}=u^{*}(u)-u^{*}_{-\infty}.\label{Nu5}
\end{eqnarray}

\begin{figure}[!ht]
\begin{center}
\includegraphics[scale=0.45]{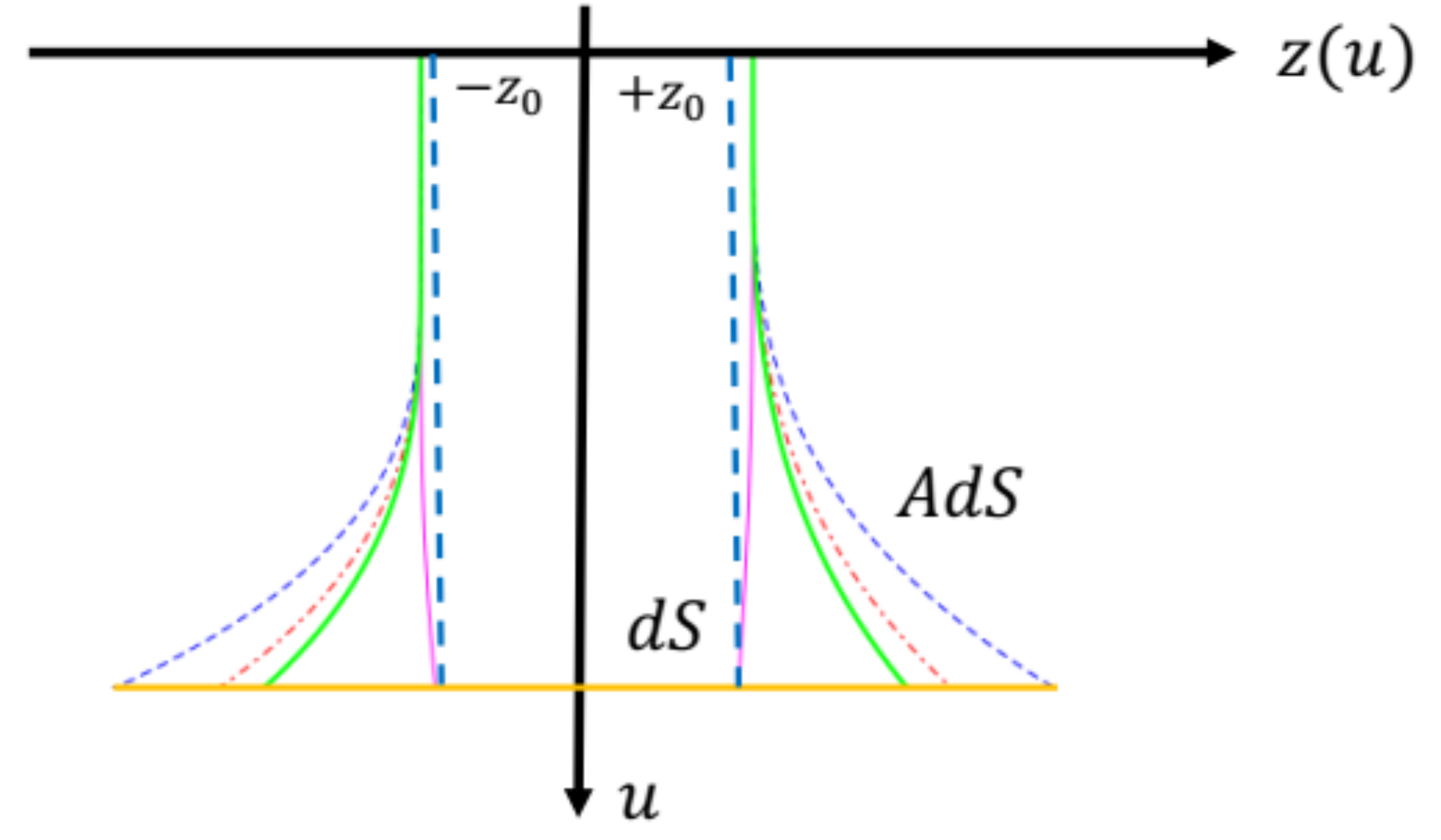}
\caption{{\it{The behavior of $z(u)$ of (\ref{Nu9}), for   
 $\Lambda=0$, $\Sigma=0.5$, $\kappa=1/4$, $\alpha=8/3$, $\gamma=0.1$ 
(solid - pink), $\gamma=1$ (dashed - blue), $\gamma=2$ (dot-dashed 
red) and $\gamma=3$ (thick - green).}
}}
\label{pfl}
\end{center}
\end{figure}

We proceed by defining the future-directed normal vectors to evaluate $K$ as
\begin{eqnarray}
n^{M}=\left(0,0,0,\frac{\dot{z}(u)f(u)}{g(u)},\frac{-1}{g(u)}\right),\label{Nu6}
\end{eqnarray}
where $g^{2}(u)=1+\dot{z}^{2}(u)f(u)e^{2A(u)}$ with the induced metric reading as
\begin{eqnarray}
ds^{2}_{ind}=e^{2A}(f(u)d\tau^{2}+dx^{2}+dy^{2})+\frac{g^{2}(u)}{f(u)}du^{2}.\label{Nu7}
\end{eqnarray}
Thus, the extrinsic curvature is given by
\[K_{\mu\nu}=\left[
\begin{array}{clcr}\label{curvature}
-\frac{e^{2A(u)}(2f(u)\dot{A}+\dot{f}(u))}{2g(u)}&0&0&0\\
0&-\frac{\dot{A}e^{2A(u)}}{g(u)}&0&0\\
0&0&-\frac{\dot{A}e^{2A(u)}}{g(u)}&0\\
0&0&0&\frac{\dot{f}(u)g(u)}{2f^{2}(u)}\\
\end{array}
\right]\]
and thus $K^{\bar{(\gamma)}}=-\frac{3\dot{A}(u)}{g(u)}$. Hence, solving equation (\ref{L10}), we find
\begin{eqnarray}
\dot{z}(u)=\frac{\Sigma}{\sqrt{4+\dfrac{\gamma\dot{\phi}^{2}}{4}-\Sigma^{2}e^{2A(u)}f(u)}}.\label{Nu9}
\end{eqnarray}
We can use (\ref{Nu9}) to draw  the regions flows, and we present them in Fig. \ref{pfl}. Note that for the fine-tuned Minkowski solution by the tension $\Sigma$ and the Horndeski-like parameters, the warp factor is just $k|u|\to\ln(\cosh(ku))$ and the graviton is marginally bound on the brane. Indeed, this is a bound state at the threshold. In this sense, the onset of the continuum is not separated by a mass gap. Besides, if we increase the tension, then the backreaction on the warp factor is even stronger, although $A(u)$ starts as linear in $u$, as in the Minkowski case. Additionally, we can see that Horndeski-like parameters change the form of the warp factor, and we acquire a transition from the dS geometry to AdS. In the dS case, $z_{0}$ is the distance between the brane and the horizon, whereas in the AdS case, $z_{0}$ is the distance to the turnaround point in the warp factor.

From the point of view of the AdS flow, which is presented in the Sec. \ref{V3}, we have that the condition of positive energy is reformulated by the well-known $c$-theorem \cite{Freedman:1999gp}, which states that $\ddot{A}\leq\,0$ \cite{Karch:2000ct}. This is one of the reasons to believe that the location of gravity on a positive stress brane is accompanied by a bending factor that asymptotically tends towards the AdS horizon, since it cannot rotate. In our solution we assume that $\ddot{A}(u)$ is positive, even though our brane has a positive stress-energy tensor, so it does not violate the positivity of the stress-energy tensor.

Let us now examine whether  this is consistent with the $c$-theorem. For the derivation of the $c$-theorem  Lorentz invariance is required, and  therefore it is only valid for Minkowski solutions. Note that  the AdS$_{4}$ brane violates $\ddot{A}\leq\,0$, since
\begin{eqnarray}
\ddot{A}(u)=-\frac{9\alpha^{2}}{64\gamma^{2}}\sech^{2}\left(\frac{3\alpha}{8\gamma}\,u\right).
\end{eqnarray}
In scenarios where gravity is located on a positive stress brane accompanied by a warp factor, asymptotically tends towards the AdS horizon  which cannot rotate. However, our solution satisfies the requirement $\ddot{A}\leq\,0$, where for $\ddot{A}\to\,0$ implies $\alpha/\gamma\to\,0$. On the other hand, we can use the fact that the energy-momentum tensor has the form
\begin{eqnarray}
&&T^{M}_{N}=diag(\rho,-p_{xx},-p_{yy},-p_{zz},-p_{uu}),\\
&&\rho=\frac{\alpha\,f\dot{\phi}^{2}}{2}+\frac{3\gamma\,f\dot{\phi}}{4}[\dot{
\phi}(3\dot{A}\dot{f}+2f(2\dot{A}^{2}+\ddot{A}))+4f\dot{A}\ddot{\phi}],\\
&&p_{xx}=\frac{\alpha\,f\dot{\phi}^{2}}{2}+\frac{\gamma\dot{\phi}}{4}[\dot{\phi}
(f(13\dot{A}\dot{f}+\ddot{f})+6f^{2}(\ddot{A}+2\dot{A}^{2})+\dot{f}^{2}
)\nonumber\\
&&+2f\ddot{\phi}(6f\dot{A}+\dot{f})],\\
&&p_{rr}=\frac{\alpha\,f\dot{\phi}^{2}}{2}-\frac{9\gamma\,f}{4}\dot{A}\dot{\phi}
^{2}(4f\dot{A}+\dot{f}),
\end{eqnarray}
where $p_{xx}=p_{yy}=p_{zz}$. The weak energy condition is $T^{\mathcal{M}}_{MN}n^{M}n^{N}\geq\,0,$
where $n^{M}$ is a null vector, or alternatively $\rho+p_{ii}\geq\,0$.  

Furthermore, we impose the null energy condition  
\begin{eqnarray}
\mathcal{S}^{\mathcal{\partial\,M}}_{\alpha\beta}n^{\alpha}n^{\beta}\geq\,
0,\label{bdry}
\end{eqnarray}
where
\begin{eqnarray}
n^{\beta}=\left(0,0,0,\frac{\dot{z}(u)f(u)}{g(u)},\frac{-1}{g(u)}\right).
\end{eqnarray}
We consider the matter stress-energy tensor on $\mathcal{\partial\,M}$, and therefore (\ref{bdry}) becomes 
equivalent to
\begin{eqnarray}
\ddot{z}(u)=-\dfrac{\Sigma\left(\frac{\gamma}{2}\dot{\phi}\ddot{\phi}-\Sigma^{2}
e^{2A(u)}(\dot{f}(u)+2f(u)\dot{A}(u))\right)}{2\left(4+\dfrac{\gamma\dot{\phi}^{
2}}{4}-\Sigma^{2}e^{2A(u)}f(u)\right)^{\frac{3}{2}}}\leq\,0.
\end{eqnarray}
Thus, $\Sigma=0$ implies $\ddot{z}=0$. We see that the dual explicit solutions for the RG flow that interpolates between dS and AdS are possible due to the variation of the parameter $\gamma$. In this sense, the Lorentzian configuration turns out to be an AdS and dS brane in the Poincar\'e path, where 
time leads to an evolution of entanglement entropy in a dS phase for small values of $\gamma$ to an AdS phase for large values of $\gamma$. Our motivation for probing this transition lies in the fact that entanglement and complexity growth for dS space must be ultra-fast \cite{Susskind:2021esx,Shaghoulian:2022fop}. This will be discussed in the following section.

\subsection{Holographic entanglement entropy in Horndeski-like gravity}\label{HEHG}

\begin{figure}[!ht]
\centerline{\includegraphics[scale=0.65]{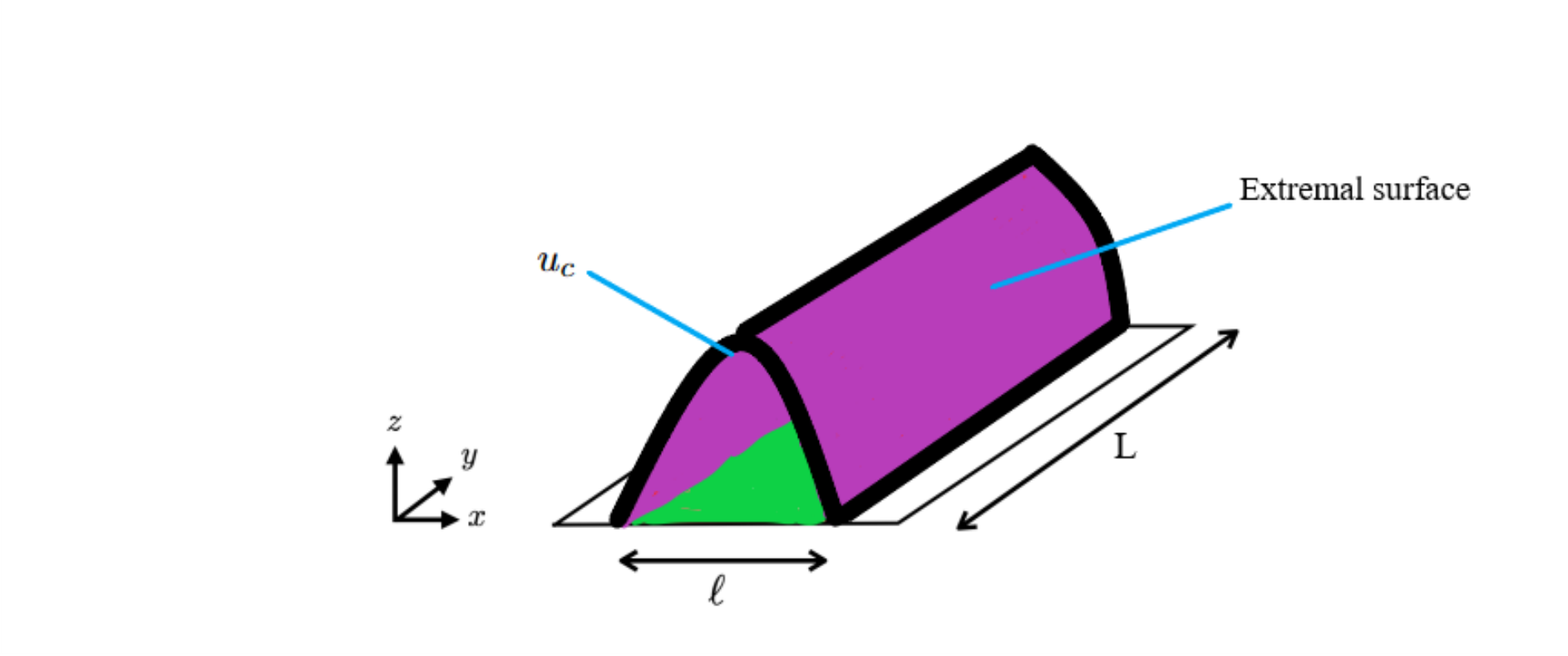}}
\caption{
{\it{Schematic representation of an extremal surface, where $l$ is the length of the subsystem $A$, which is anchored on the subsystem living on the boundary.}
}}\label{SCHEM}
\label{planohwkhz}
\end{figure}

In this section, we present the computations of the entanglement entropy to 
enable the growth process of the information between the AdS and dS phase in 
Horndeski-like gravity, following the procedures of 
\cite{Fabiano-JHAP,Takayanagi:2012kg,Calabrese:2004eu,Ryu:2006ef,Ryu:2006bv,
Susskind:1994sm}. We consider the metric (\ref{metricflat}) where 
the four-dimensional CFT lives in the space measured by $t$ and $x$. In this 
sense we can choose the subsystem $A$, having length $l$ in the interval 
$x\epsilon[-l/2,l/2]$, $y\epsilon[L/2,L/2]$, $z\epsilon[L/2,L/2]$, as it is 
shown in Fig. \ref{SCHEM}. 

As shown in \cite{Fabiano-JHAP}, the entanglement entropy for the Horndeski-like gravity has a correction to Ryu-Takayanagi formula \cite{Ryu:2006ef,Ryu:2006bv}. Motivated by recent studies of \cite{Santos:2021orr}, and following the steps of \cite{Caceres:2017lbr,Hajian:2020dcq} through the induced metric ($u_{c}$ is a constant of integration representing the turning point of the extremal surface in the higher dimensional AdS$_5$ bulk spacetime (see Fig. \ref{SCHEM}))
\begin{eqnarray}
ds_{ind}=\frac{u^{2}_{c}du}{u^{3}\sqrt{\left(1-\frac{u^{4}_{c}}{u^{4}}
\right)f(u)}},\label{T5}
\end{eqnarray} 
we can provide the Ryu-Takayanagi formula for Horndeski-like gravity as
\begin{eqnarray}
&&S_{A}=\frac{\mathcal{A}}{4G_{N}},\label{ES01}\\
&&\mathcal{A}=\int{ds_{ind}\Phi},\\
&&\Phi=1-2\gamma(\bar{\gamma}^{\lambda\sigma}\nabla_{\lambda}\phi\nabla_{\sigma}
\phi).
\end{eqnarray}
Note that the area integral is divergent at the point $u=u_{c}$ and must be regularized introducing an infrared cutoff ($u_{b}$). Nevertheless, using the holographic dictionary, we have a relation between the UV cutoff of the boundary field theory ($\epsilon$) and to bulk IR cutoff. Such relation is 
inversely related through the AdS length scale $\mathcal{R}$ and can be established as  $u_{b}=\mathcal{R}/\epsilon$. Furthermore, beyond the area integral, we can obtain the length of the subsystem as
\begin{eqnarray}
\frac{l}{2}=\int^{0}_{u_{c}}{\frac{du}{\sqrt{\left(1-\frac{u^{4}}{u^{4}_{c}}
\right)e^{2A(u)}f(u)}}}.\label{ES2} 
\end{eqnarray}

Now, we are ready to investigate the behavior of the entanglement entropy that interpolates between AdS and dS regimes.  

\subsection{Case A}

We start with case A of subsection \ref{case1}, namely the exponential superpotential, where the entanglement entropy is obtained as
\begin{eqnarray}
S_{A}=\frac{1}{4G_{N}}\int^{\epsilon}_{u_{c}}{\frac{u^{2}_{c}\Phi\,du}{u^{3}
\sqrt{\left(1-\frac{u^{4}_{c}}{u^{4}}\right)f(u)}}},\quad\,f(u)=\dfrac{3e^{
3\sqrt{\frac{8}{3}}\ln(u/u_{c})}}{16\left(\sigma-\frac{2}{\gamma}\right)},
\label{entan}.
\end{eqnarray} 
Far away from the extremal surface, namely $u_{h}<<u_{c}$, we can perform Taylor expansion, obtaining   
 \begin{eqnarray}
S_{A}=\frac{1}{4G_{N}}\left(\frac{1}{2\sqrt{f(0)}}+\frac{27\gamma^2}{64e^{2A(0)}
f^{3/2}(0)\left(2\sqrt{6}-4\right)(\alpha-2)l^2}\right).
\end{eqnarray}
The last divergence that we need to remove is around $f(0)$ and $A(0)$. For this, we consider $f(0)\to\,f(1-\epsilon)$ and $A(0)\to\,A(1-\epsilon)$ when $\epsilon\to\,1$, and thus we recover the usual result. Hence, through the Taylor series, we have
\begin{eqnarray}
S_{A}=\frac{8}{3G_{N}}\left[\frac{\alpha-2}{2\gamma}+\frac{27\sqrt{
2\gamma(\alpha-2)}}{16\sqrt{3}\left(2\sqrt{6}-4\right)l^2}\right].
\end{eqnarray}
For small values of $\gamma$ the dS phase becomes dominant, while for high values of $\gamma$ it is the AdS phase that dominates (see Fig. \ref{pfl}). In fact, as it was shown in Fig. \ref{p3}, at $\gamma=0.1$  
we have a maximum value for the dS phase and a minimum value for the AdS phase. 

\subsection{Case B}
In the case B of subsection  \ref{case2} of vacuum solution, the entanglement entropy can be obtained as
\begin{eqnarray}
S_{A}=\frac{1}{4G_{N}}\int^{\epsilon}_{u_{c}}{\frac{u^{2}_{c}\Phi\,du}{u^{3}
\sqrt{\left(1-\frac{u^{4}_{c}}{u^{4}}\right)f(u)}}},\quad\,f(u)=\frac{3\sigma}{
4W^{2}_{0}}.\label{entan1}
\end{eqnarray}
Following the same steps as the above, we have 
 \begin{eqnarray}
S_{A}=\frac{1}{4G_{N}}\left\{\sqrt{\frac{\alpha+\gamma-1 
}{\gamma(\alpha-1)}}+\frac{27\gamma^2}{8\left(2\sqrt{6}-4\right)(\alpha-2)l^2}
\left[\frac{\alpha+\gamma-1 }{\gamma(\alpha-1)}\right]^{3/2}\right\},
\end{eqnarray} 
where for $\alpha>2$ no singularities appear. Similarly to before, in small values of $\gamma$ the dS phase become dominant, while for large values of $\gamma$ it is the AdS phase that dominates.

\subsection{Case C}

In the case C of the smooth solution of subsection \ref{case3}, the entanglement entropy is obtained as
\begin{eqnarray}
S_{A}&=&\frac{1}{4G_{N}}\int^{\epsilon}_{u_{c}}{\frac{u^{2}_{c}\Phi\,du}{u^{3}
\sqrt{\left(1-\frac{u^{4}_{c}}{u^{4}}\right)f(u)}}},\nonumber\\
f(u)&=&\frac{3 \left(15 k^4 \text{sech}^4(k u)-12 k^4 \text{sech}^2(k u)-8 k^2 
\text{sech}^2(k u)+8 k^2\right)}{2 
\left(\sigma-\frac{2}{\gamma}\right)}.\label{entan2}
\end{eqnarray}
After the expansions we find 
\begin{eqnarray}
S_{A}=\frac{1}{4G_{N}}\left\{\frac{1}{2}\sqrt{\frac{32(\alpha-1)}{27\alpha}}
+\frac{27\gamma^2}{64\left(2\sqrt{6}-4\right)(\alpha-2)l^2}\left[\frac{
32(\alpha-1)}{27\alpha}\right]^{3/2}\right\}.
\end{eqnarray}
The entanglement entropy for $\alpha=1$ has no dS and AdS phases. The limit 
$\gamma\to0$  represent a dS phase, with  
\begin{eqnarray}
&&S^{dS}_{A}=\frac{\mathcal{A}}{4G_{N}},\\
&&\mathcal{A}=\frac{1}{2}\sqrt{\frac{32(\alpha-1)}{27\alpha}},
\end{eqnarray}
where we have Randall-Sundrum dS brane \cite{Santos:2021orr,Santos:2023flb}. Note that for $\gamma=0.1$ we have a maximum value for the dS phase and a minimum value for the AdS phase (see Fig. \ref{p3}). Hence, we obtain two entropy parts, i.e. $S_{A}=S^{dS}_{A}+S^{AdS}_{A}$ with 
\begin{eqnarray}
S^{AdS}_{A}=\frac{1}{4G_{N}}\left\{\frac{27\gamma^2}{64\left(2\sqrt{6}
-4\right)(\alpha-2)l^2}\left[\frac{32(\alpha-1)}{27\alpha}\right]^{3/2}\right\}.
\end{eqnarray}
An interesting aspect of this case is that dS part has no $\gamma$-dependence, while in case A of above the two parts of the entropy did have a $\gamma$-dependence.

\section{RG flow equations}
\label{V3}

In this section we investigate the holographic scenario through the warp factor of a spacetime geometry \cite{Bazeia:2006ef}, for which the five-dimensional metric is
\begin{eqnarray}
ds^{2}=\frac{du^{2}}{f(u)}+a^{2}(u)[-f(u)dt^{2}+dx^{2}+dy^{2}+dz^{2}],
\label{RG.1}
\end{eqnarray}
where $a(u)=e^{A(u)}$. In the domain wall/QFT correspondence 
\cite{Boonstra:1998mp,Freedman:1999gp,linde,cvetic,LopesCardoso:2002ec,
Bazeia:2006ef} the warp factor is identified with the renormalization scale 
$a(u)$ of the flow equations.

We consider multi-running couplings $\phi^{i}$. Thus, from (\ref{RG.1}) we have
\begin{eqnarray}
\dot{\phi}^{i}(u)=\frac{d\phi^{i}}{du}=\frac{da}{du}\frac{d\phi^{i}}{da}=\dot{A}
a\frac{d\phi^{i}}{da},\quad\phi^{i}=(\phi,\chi).\label{RG.2}
\end{eqnarray}
According to the aforementioned correspondence, the scalar field on the gravity side is conjectured to be related to the running coupling on the dual field theory side \cite{Kiritsis:2014kua,Anninos:2017hhn,Kiritsis:2016kog}. Furthermore, we can construct a beta ($\beta(\phi^{i})$) function of the boundary QFT in terms of $\phi^{i}$ as 
\begin{eqnarray}
\frac{d\phi^{i}}{d\log\mu}=\beta(\phi^{i})\equiv 
a\frac{d\phi^{i}}{da}=\frac{\dot{\phi}^{i}}{\dot{A}}=-\frac{3}{2}\frac{W_{\phi^{
i}}(\phi^{i})}{W(\phi^{i})},\label{RG.3}
\end{eqnarray}
where $\mu\equiv\mu_{0}e^{A(u)}$ is the (dual) QFT energy scale and $\mu_{0}$ is an arbitrary mass scale.  

For critical points $\phi^{i}=\phi^{i*}$ (or $\phi^{i}=\phi^{i}_{vac}$ for supersymmetric vacua) the $\beta(\phi^{i})$ function vanishes. Performing an expansion of the $\beta(\phi^{i})$-function around the critical points we find
\begin{eqnarray}
\beta(\phi^{i})=\beta(\phi^{i*})+\beta^{'}(\phi^{i*})(\phi^{i}-\phi^{i*})+...,
\label{RG.4}
\end{eqnarray}
where
\begin{eqnarray}
\beta^{'}(\phi^{i})=\frac{3}{2}\left(-\frac{W_{\phi^{i}\phi^{i}}}{W}+\frac{W^{2}
_{\phi^{i}}}{W^{2}}\right)_{\phi^{i}=\phi^{i*}}.\label{RG.5}
\end{eqnarray}
Note that combining (\ref{RG.3}) and (\ref{RG.4}), and integrating out both sides with $\beta(\phi^{i*})=0$, one can find the running coupling equation
\begin{eqnarray}
\phi^{i}=\phi^{i*}+sa^{\beta^{'}(\phi^{i*})},\label{RG.6}
\end{eqnarray}
where $s$ is a constant. The regime $\beta^{'}(\phi^{i*})<0$ and energy scale $a\to\infty$ implies an ultraviolet (UV) stable fixed point, whereas for $\beta^{'}(\phi^{i*})>0$ and energy scale $a\to 0$ we have an infrared (IR) stable fixed point. 

\subsection{Case A}

\begin{figure}[h]
\centerline{\hspace{-3cm}\includegraphics[scale=0.6]{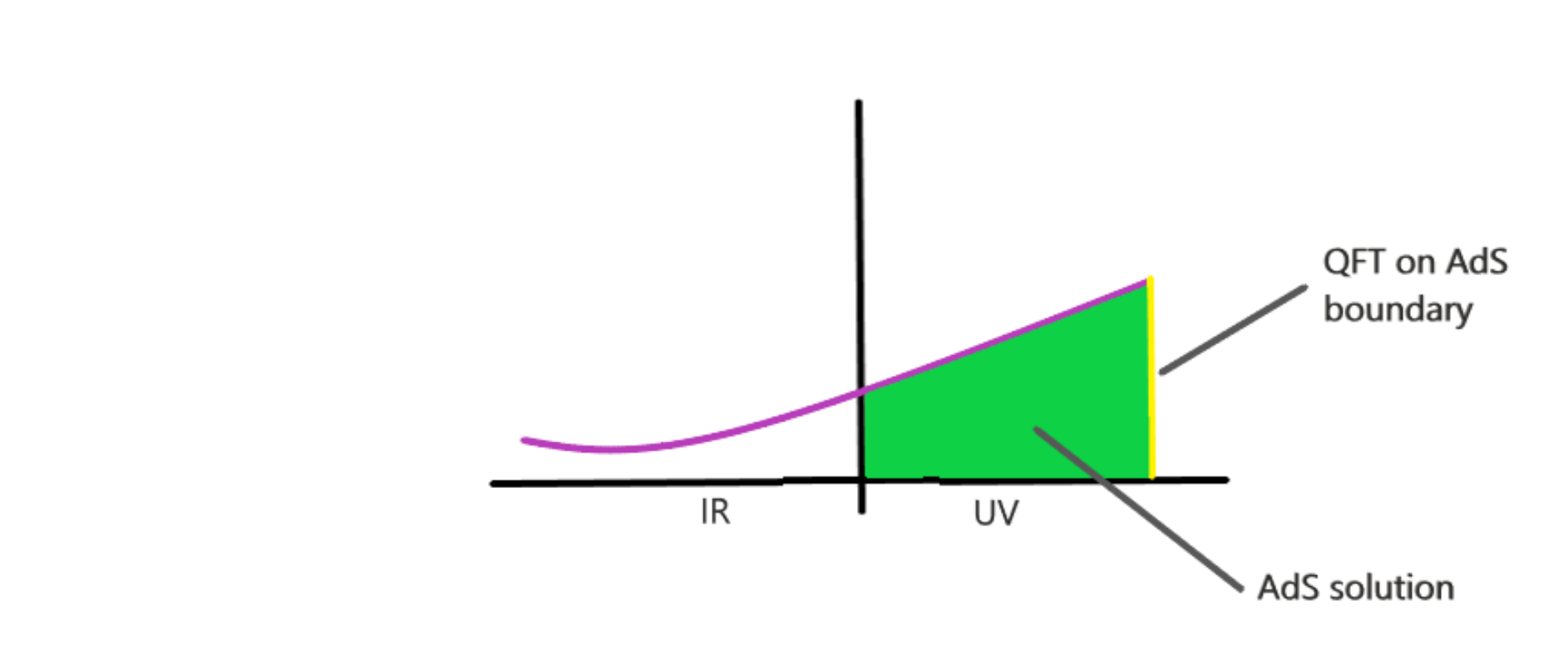}}
\caption{{\it{Schematic representation of the RG flow  for case A of   
  exponential superpotential, with a stable fixed point in the UV 
regime.}}}
\label{UVRE}
\end{figure}

In the case A of subsection \ref{case1} of exponential superpotential, inserting $W(\phi^{i})=e^{\sqrt{\frac{8}{3}}\phi^{i}}$ into equation (\ref{RG.5}), we acquire $\beta^{'}(\phi^{*})=0$, and thus see that $a(u)=u^{1/4}$. Note that $u\to\infty$ for $a\to\infty$, and this regime 
is the ultraviolet (UV) stable fixed point. In  Fig. \ref{UVRE} we present a schematic representation of RG flow, and we see that in the quantum field theory frame of reference, the charge $a_{UV}$ at the UV fixed point is greater than $a_{IR} $ of the IR fixed point.

\subsection{Case B}

\begin{figure}[!ht]
\centerline{\hspace{-1cm}\includegraphics[scale=0.6]{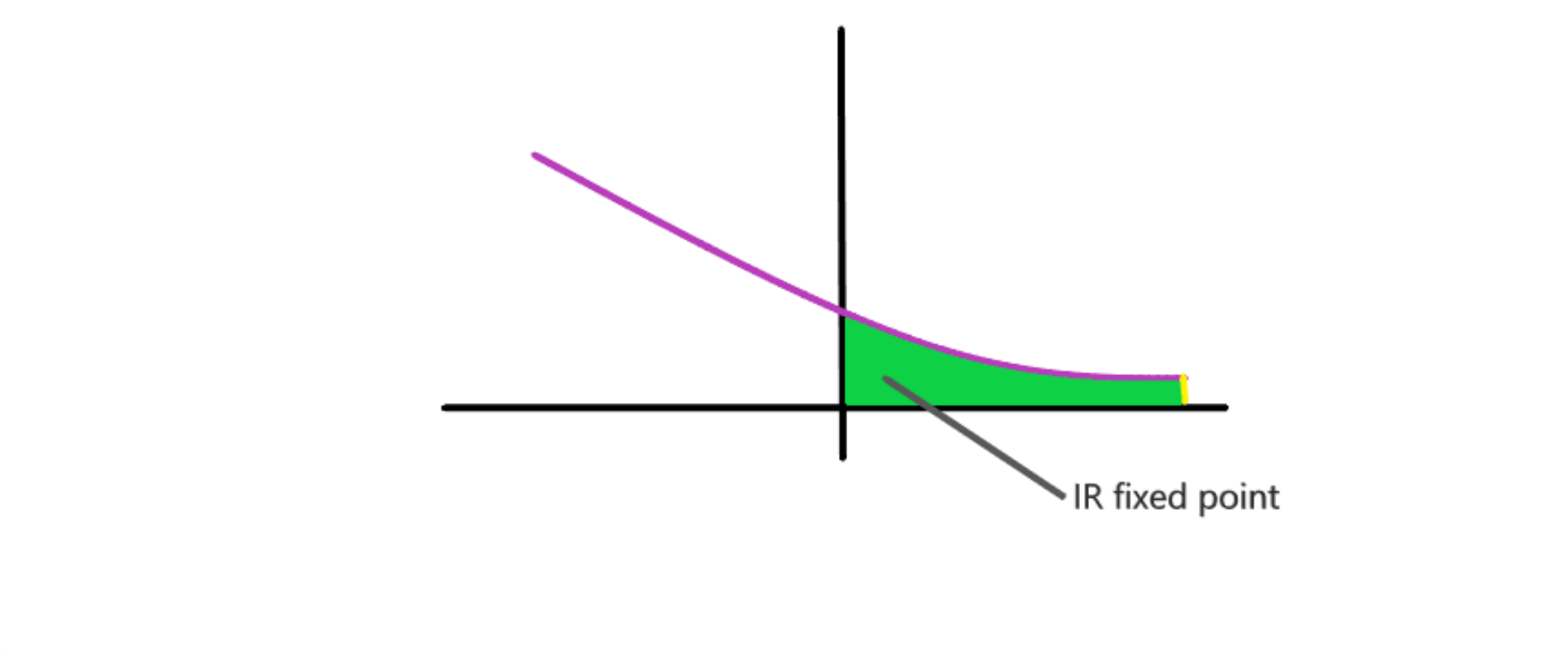}}
\caption{{{Schematic representation of RG flow for case B of   vacuum solution, 
  with a stable fixed point in the IR regime.}}}
\label{UVRE1}
\end{figure}

In the case B of subsection \ref{case2} of vacuum solution, inserting the superpotential $W_{0}=\sqrt{3(\sigma+2/\gamma)}/4$ into equation (\ref{RG.5}), we have $\beta^{'}(\phi^{*})=0$. Hence,  we have that $\phi,\chi=const.$ and the warp  factor can be found as $A=-(1/3)W_0\,u$, therefore 
$a=e^{-(1/3)W_0\,u}$. This regime is the  IR stable fixed point. In Fig. \ref{UVRE1} we present the schematic representation of the corresponding RG flow, where we observe that now $a_{UV}$  is smaller and $a_{IR}$ is larger. Hence, the RG flow is a decreasing function, and this is a holographic proof of $c$-theorems that have been obtained in \cite{Freedman:1999gp}.

\subsection{Case C}

\begin{figure}[!ht]
\centerline{\includegraphics[scale=0.6]{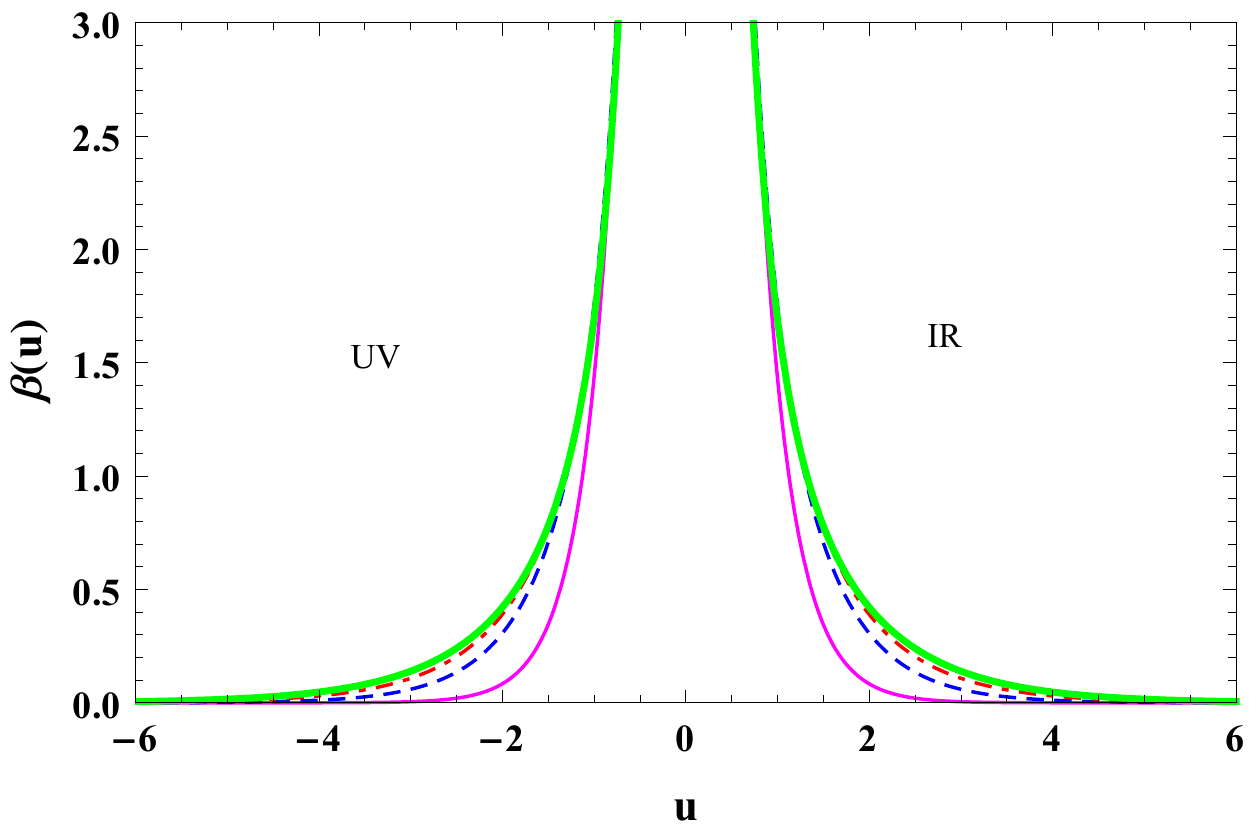}}
\caption{{\it{
The 
behavior of $\beta(u)$ according to  (\ref{RG.7}), for   
$\Lambda=0$, $\kappa=1/4$, $\alpha=8/3$, $\gamma=0.1$ (solid - pink), 
$\gamma=1$ (dashed - blue), $\gamma=2$ (dot-dashed - red) and 
$\gamma=3$ (thick - green).}}}
\label{UVRE2}
\end{figure}

In the case C of the smooth solution of subsection \ref{case3}, for the superpotential $W(u)=3k\tanh(ku)$ we obtain
\begin{eqnarray}
&&\beta^{'}(u)=\frac{3k^{2}}{2}[(csch^2(k u)+2)\sech^2(k 
u)]_{u=u^{*}},\label{RG.7}\\
&&\beta^{'}(u)=\frac{12 k^2}{\left(e^{-k u}+e^{k u}\right)^2}+\frac{24 
k^2}{\left(e^{k u}-e^{-k u}\right)^2 \left(e^{-k u}+e^{k u}\right)^2},\\
&&k=\frac{1}{2}\sqrt{\frac{3\alpha}{2\gamma}}.
\end{eqnarray}
For this case $a=e^{-k\tanh(ku)}$, which provides that the charge $a_{UV}$ at the UV fixed point is smaller and $a_{IR}$ at the IR fixed point is larger, being an increasing function. This can be seen in Fig. \ref{UVRE2}, where we present the behavior of $\beta(u)$ according to (\ref{RG.7}).

\subsection{Tensor perturbations }\label{tensorpertur}

In this section, we proceed to the examination of tensor perturbations, in order to study the gravity localization on the black hole ansatz with flat slicing, by considering the new superpotential solutions. Considering $\eta_{\mu\nu}+\epsilon\,h_{\mu\nu}$, we have
\begin{eqnarray}
ds^{2}=\frac{du^{2}}{f(u)}+e^{2A(u)}(\eta_{\mu\nu}+\epsilon\,h_{\mu\nu})dx^{\mu}
dx^{\nu},\label{tensor}
\end{eqnarray}
where $\delta^{(1)}g_{\mu\nu}=h_{\mu\nu}$ is the 
 first-order perturbation   being transverse and traceless (TT), namely  
$\eta^{\mu\alpha}=\partial_{\alpha}h_{\mu\nu}=0$ and 
$h\equiv\eta^{\mu\nu}h_{\mu\nu}=0$ \cite{Brito:2018pwe,Santos:2021guj}. Then we 
can write 
\begin{equation}
T(u)\ddot{h}_{\mu\nu}+B(u)\dot{h}_{\mu\nu}-e^{-2A(u)}\Box_{4d}h_{\mu\nu}=0,
\label{tensor1}
\end{equation}
where
\begin{eqnarray}
B(u)&=&\frac{4\dot{A}(u)-4\gamma\,f(u)\dot{A}(u)\dot{\phi}^{2}-\dot{f}
(u)/2f(u)-3\gamma\dot{f}\dot{\phi}^{2}/2-2\gamma\,f(u)\dot{\phi}(u)\ddot{\phi}
(u)}{1+\gamma\,f(u)\dot{\phi}^{2}(u)},\nonumber\\
T(u)&=&\frac{1-\gamma\,f(u)\dot{\phi}^{2}(u)}{1+\gamma\,f(u)\dot{\phi}^{2}(u)}.
\end{eqnarray}
Note that for $f(u)\to1$  we recover the usual results of 
\cite{Brito:2018pwe,Santos:2021guj}, while for $\gamma\to\,0$ together with
$f(u)\to1$,   equation (\ref{tensor1}) recovers   the usual form of 
Karch-Randall one (see \cite{Karch:2000ct,Randall:1999vf} for more 
discussions).

We remind that, as we showed in section \ref{V3} above, only cases B and C are capable of localizing gravity. Thus, in the following, we restrict our analysis in these cases.

\subsection{Case B}

\begin{figure}[!ht]
\centerline{\includegraphics[scale=0.7]{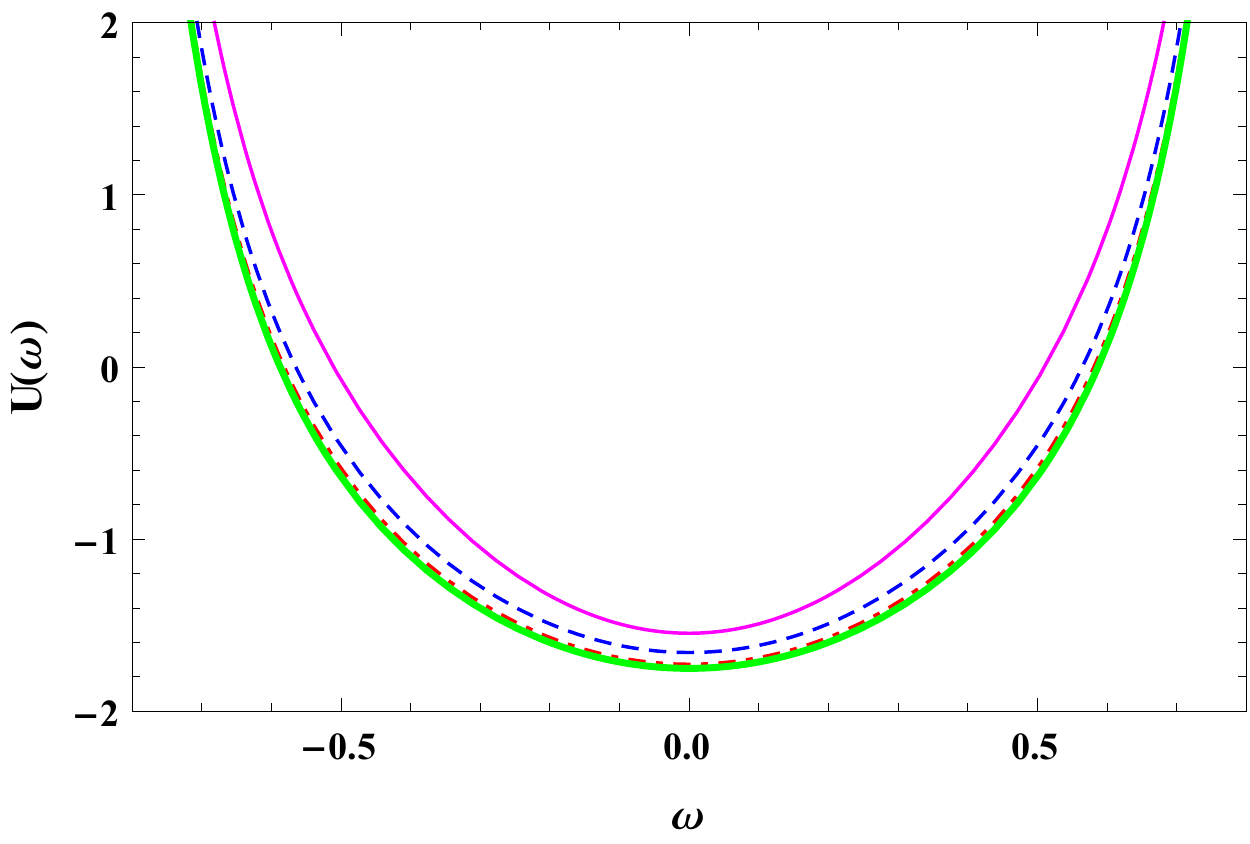}}
\caption{{\it{The box potential.}}}
\label{UVRE3}
\end{figure}

In the case B of subsection \ref{case2} of vacuum solution, for the superpotential $W_{0}=\sqrt{3(\sigma+2/\gamma)}/4$ we have $A(u)=-(1/3)W_0\,u$ and $f(u)=3\sigma/4W^{2}_{0}$. Inserting these into equation (\ref{tensor1}) the coefficients become $T(u)=1$ and $B(u)=4\dot{A}(u)$, and therefore
\begin{eqnarray}
\ddot{h}_{\mu\nu}+4\dot{A}(u)\dot{h}_{\mu\nu}-e^{-2A(u)}\Box_{4d}h_{\mu\nu}=0.
\label{tensor2}
\end{eqnarray}
Considering coordinate transformation  $du=e^{A}d\omega$, and imposing the decomposition 
$h_{\mu\nu}(x,\omega)=\mathcal{E}_{\mu\nu}e^{-ipx}e^{-3A/4}H(\omega)$ with 
$p^{2}=-m^{2}$, this equation simplifies to 
\begin{eqnarray}
&&-\partial^{2}_{\omega}H(\omega)=E^{2}_{m}H(\omega),\label{tensor4}\\
&&H_{m}(\omega)=\cos(E_m\omega),\\
&&E^{2}_{m}=m^{2}-V(\omega),
\end{eqnarray}
where $V(\omega)=3\alpha/8\gamma$ is a constant, while for the usual value of $\alpha=8/3$ we find $V(\omega)=1/\gamma$. This potential gives rise to a toy model, which is adequate  to calculate the Kaluza-Klein (KK) modes exactly \cite{Randall:1999vf,Csaki:2000fc}. In particular, it resembles the known potential of the volcano box (see Fig. \ref{UVRE3}), which is zero in the regions $\omega>\omega_1$, $\omega>\omega_2$, where there cannot be a limit state with zero energy. Additionally, one can suitably arrange the depth and width of the well so that there is a single bound-state, with small vanishing energy $m^{2}<0$, and a continuum for $m^{2}\geq\,0$. The energy $m^{2}=E^{2}_{m}+V(\omega)$ is a state that lies above the square well.

Now, the correction of any continuous states $H_{n}(\omega)$ can be 
obtained by integrating such states for which the state density measure is 
relevant. Thus, the correction of Newton's Law is
\begin{eqnarray}
V(\omega)=H^{2}_{0}(0)\frac{e^{-m\omega}}{\omega}+\int{dm\,m^{2}\,H^{2}_{m}
(0)\frac{e^{-m\omega}}{\omega}}.
\end{eqnarray}
Based on the above expression, for decoupled continuous modes the probability 
of continuous modes tunneling in the central region of the potential is 
extremely small for $m\to\,0$:
\begin{eqnarray}
V(\omega)=e^{-m\omega}\left(\frac{1}{\omega}-\frac{2+2m\omega+ 
m^2\omega^2}{\omega^4}\right).
\end{eqnarray}


\subsection{Case C}

As mentioned above, in the case C of the smooth solution of subsection \ref{case3} we have $A(u)=-\ln{\cosh{ku}}\approx-\frac{k^2}{2}u^{2}$, while the superpotential and scalar potential are $W(u)=3ku$ and $\phi(u)=k^{2}u$ with $k=\sqrt{3\alpha/2\gamma}/2$. In the harmonic oscillator approximation \cite{Llatas:2001jj} $f(u)$ is given by
\begin{equation}
f(u)\sim\frac{\alpha}{\gamma}\left(1-\frac{u^{2}}{u^{2}_{c}}\right).
\end{equation} 
Now, following the steps of \cite{Brito:2018pwe} with $du=e^{-A}dr$, we have
\begin{eqnarray}
ds^{2}=e^{2A(r)}\left[\frac{dr^{2}}{f(r)}+(\eta_{\mu\nu}+\epsilon\,h_{\mu\nu}
)dx^{\mu}dx^{\nu}\right],\label{tensor}
\end{eqnarray}
and hence the transverse and traceless tensor perturbations follow
\begin{equation}
C(r)\ddot{h}_{\mu\nu}+D(r)\dot{h}_{\mu\nu}+\Box_{4d}h_{\mu\nu}=0,\label{T1}
\end{equation}
where
{\small{
\begin{eqnarray}
D(r)&=&\frac{3\dot{A}(r)-\gamma\,f(r)e^{-2A(r)}\dot{A}(r)\dot{\phi}^{2}-\dot{f}
(r)/2f(r)-3\gamma\dot{f}e^{-2A(r)}\dot{\phi}^{2}/2-2\gamma\,f(r)e^{-2A(r)}\dot{
\phi}(r)\ddot{\phi}(r)}{1+\gamma\,f(r)e^{-2A(r)}\dot{\phi}^{2}(r)},\nonumber\\
C(r)&=&\frac{1-\gamma\,f(r)e^{-2A(r)}\dot{\phi}^{2}(r)}{1+\gamma\,f(r)e^{-2A(r)}
\dot{\phi}^{2}(r)}.
\end{eqnarray}}}
Performing the coordinate transformation $dr=\sqrt{C}d\omega$, and imposing the decomposition $h_{\mu\nu}(x,\omega)=\epsilon_{\mu\nu}(x)e^{-ipx}H(\omega)$ with $p^{2}=-m^{2}$, this equations simplifies to
\begin{eqnarray}
&&\partial^{2}_{\omega}H(\omega)+Q(\omega)\partial_{\omega}H(\omega)+m^{2}
H(\omega)=0,\label{T5}\\
&&Q(\omega)=\frac{D}{\sqrt{C}}-\frac{\partial_{\omega}C}{2C}\label{T6}.
\end{eqnarray}
Finally, considering $H(\omega)=G(\omega)\psi(\omega)$ with $G(\omega)=\exp\left(-\frac{1}{2}\int{Q(\omega)d\omega}\right)$, it is further transformed to a Schrödinger-like equation as:
\begin{eqnarray}
&&-\partial^{2}_{\omega}\psi(\omega)+U(\omega)\psi(\omega)=m^{2}\psi(\omega),
\label{T7}\\
&&U(\omega)=\frac{Q^{2}}{4}+\frac{\partial_{\omega}Q}{2}.\label{T8}
\end{eqnarray}

\begin{figure}[!ht]
\centerline{\includegraphics[scale=0.8]{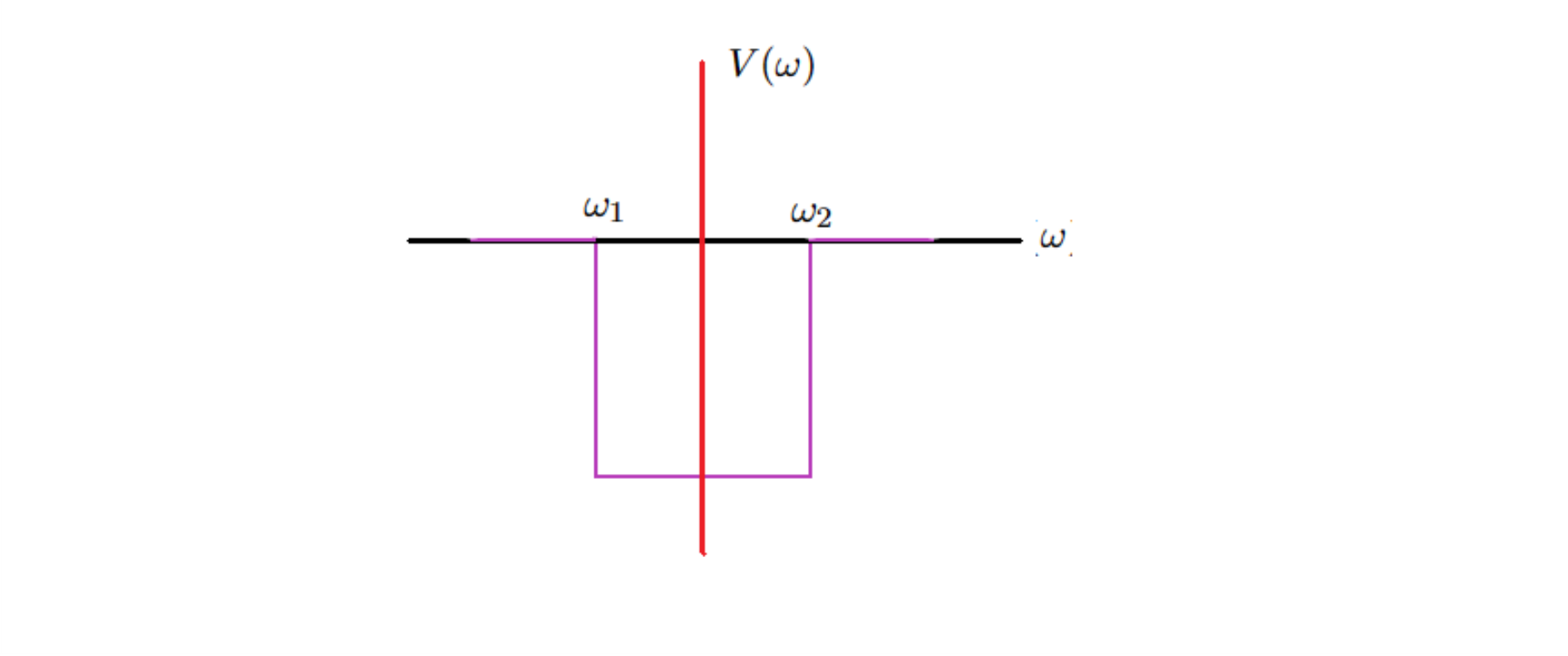}}
\caption{{\it{The potential according to (\ref{T8}), for   $\Lambda=0$, 
$\kappa=1/4$, $\alpha=8/3$, $\gamma=0.1$ (solid - pink), $\gamma=1$ (dashed 
 - blue), $\gamma=2$ (dashed-dotted - red) and $\gamma=3$ (thick - 
green), respectively.}}}
\label{UVRE3bb}
\end{figure}

In Fig. \ref{UVRE3bb} we draw the above potential, which presents an unusual profile comparing to those of the literature \cite{Csaki:2000fc,Karch:2000ct}. This behavior is of the volcano potential type, and can interpolate between asymptotically AdS spacetimes  and asymptotically dS spacetimes. 

We mention here that according to \cite{Susskind:2022bia} there are two types of limitations in working with a flat-space approximation in dS space. The first is related to  the restriction of the amount of mass, which must be small enough in order not to cause a global reaction in the geometry. The second 
requires that the energy should be large enough in order for the corresponding wavelength to be smaller than the dS scale. Thus, our black hole ansatz with flat slicing is a good approximation of flat space, and provides a good region between these limits.

\begin{figure}[!ht]
\centerline{\includegraphics[scale=0.55]{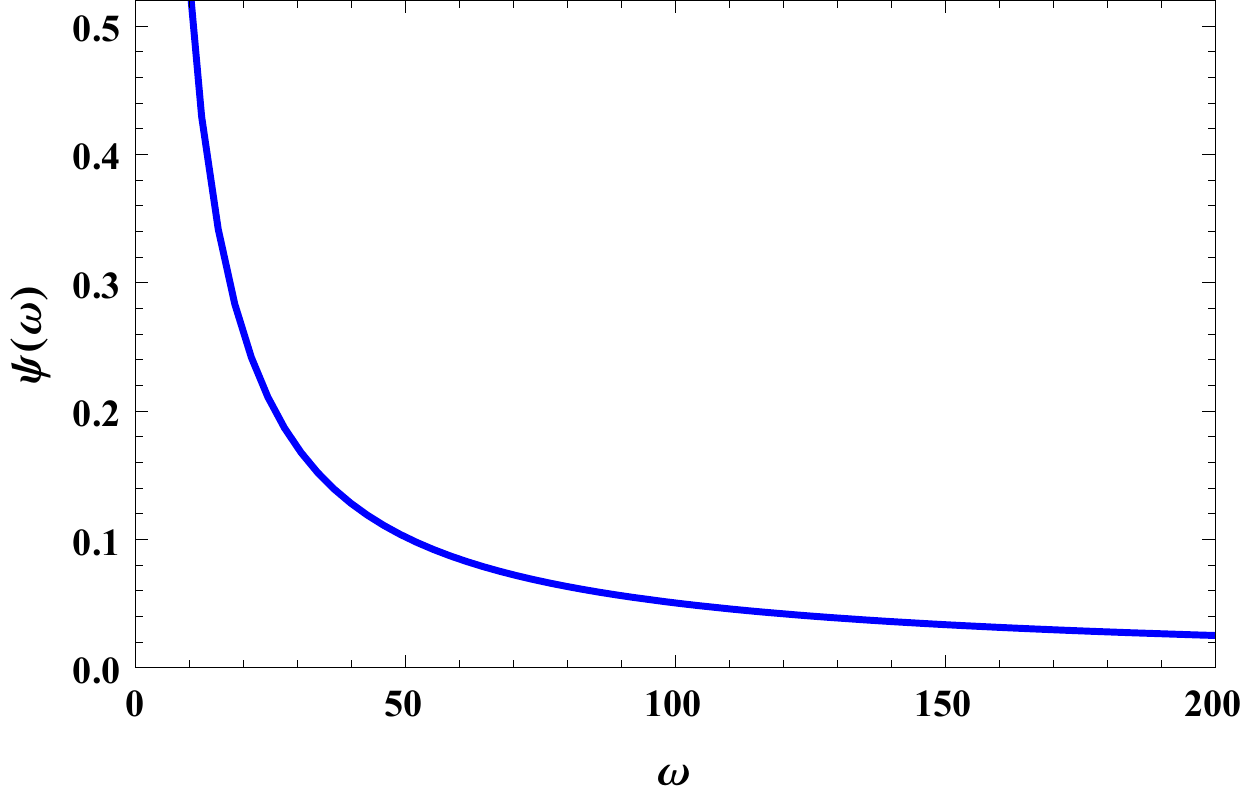}}
\centerline{\includegraphics[scale=0.55]{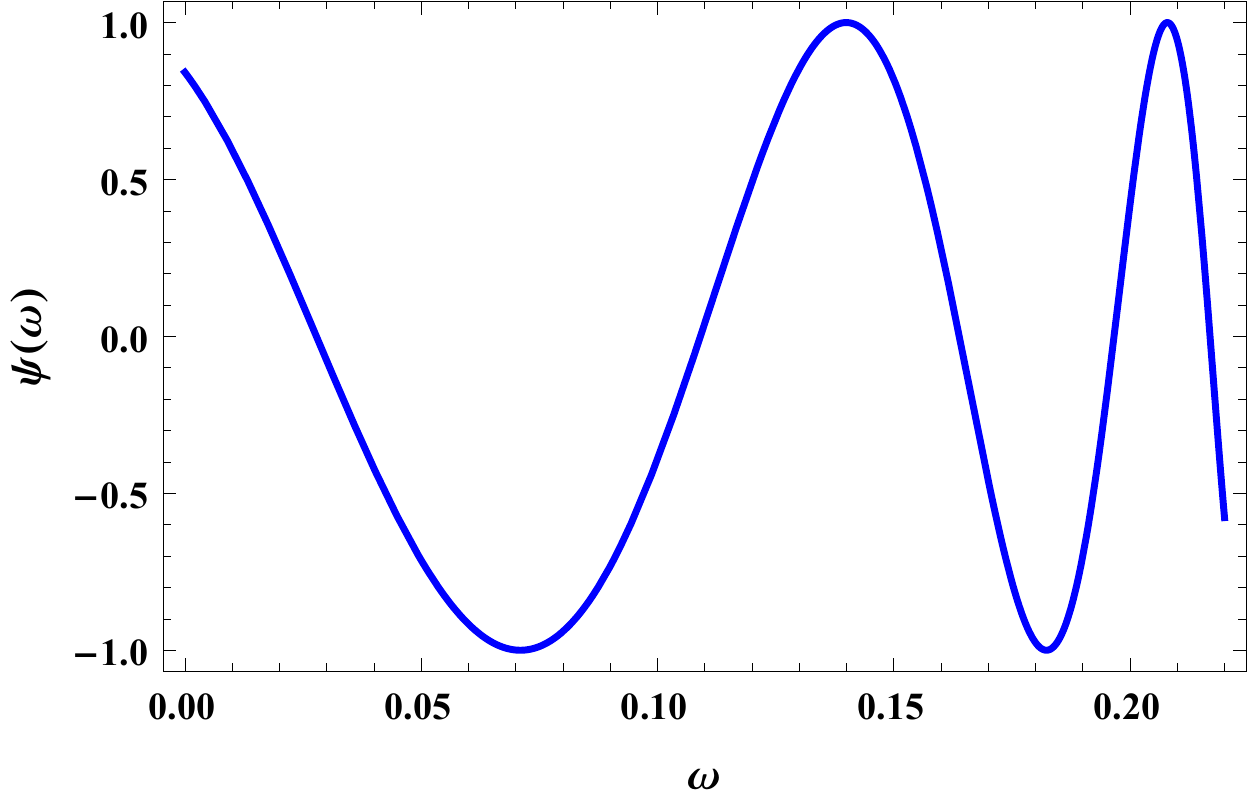}}
\caption{{\it{Top panel: the wavefunction $\psi(\omega)$ of the almost 
massless mode for the values $m^2=0.0419$ and $\gamma=0.1$. Bottom panel: 
 the wavefunction of the  highly excited mode with $m^2=6$ and 
$\gamma=2$.}}}
\label{UVRE5}
\end{figure}

We proceed to the examination of the crucial characteristics of the spectrum, obtained through a numerical solution of the Schrödinger-like equation (\ref{T7}) using the method employed in \cite{Brito:2018pwe}. In Fig. \ref{UVRE5}, we depict the wavefunctions $\psi(\omega)$ of the almost massless mode and of a highly excited mode. As expected, for $\gamma$ small and $m^{2}$ the wavefunction agrees with that obtained in \cite{Karch:2000ct}. In particular, in the flat slicing limit, the top panel approaches the zero modes located on the brane, while the bottom panel approaches a highly excited Kaluza-Klein KK mode. Additionally, the low KK modes also appear as we expect: oscillating in volume but suppressed in the brane, which can be dS or AdS. Hence, we have a dS brane in the pode and a symmetric brane in the antipode.

\begin{figure}[!ht]
\centerline{\includegraphics[scale=0.55]{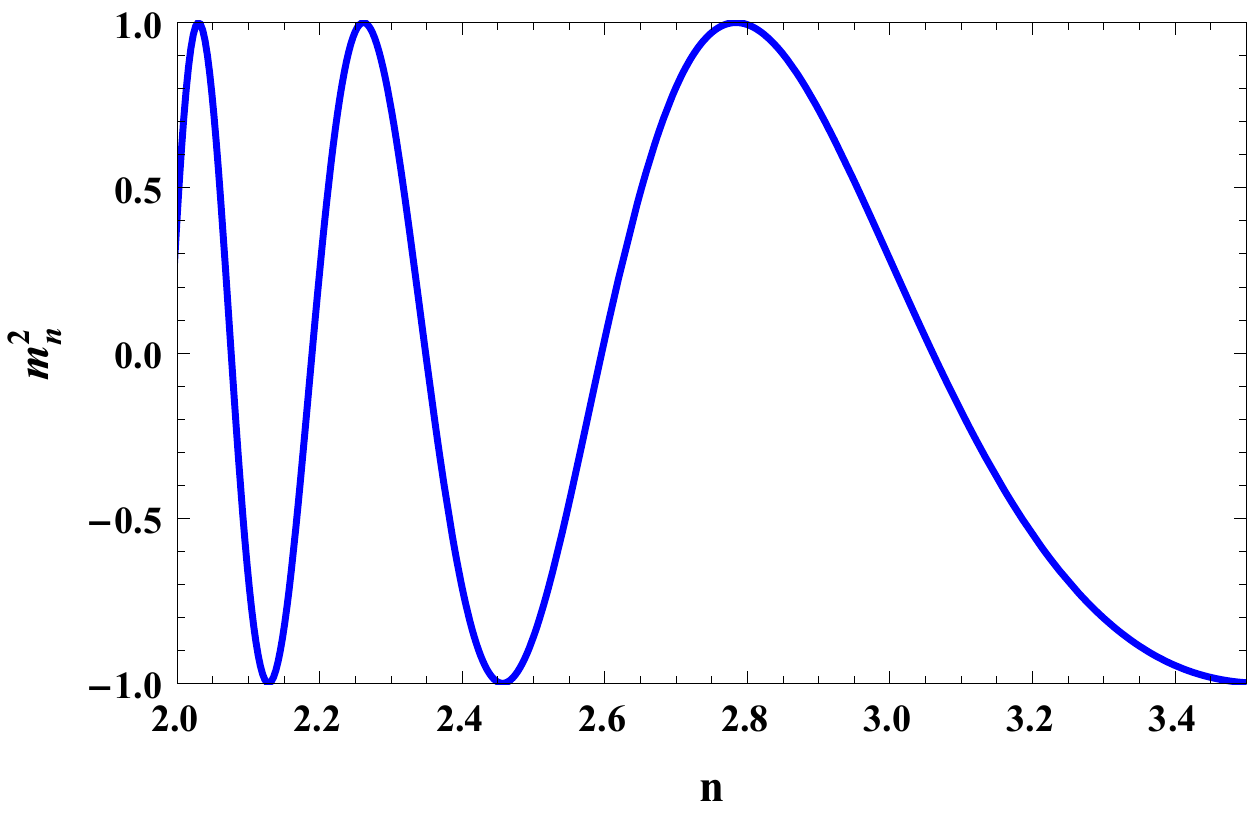}}
\centerline{\includegraphics[scale=0.55]{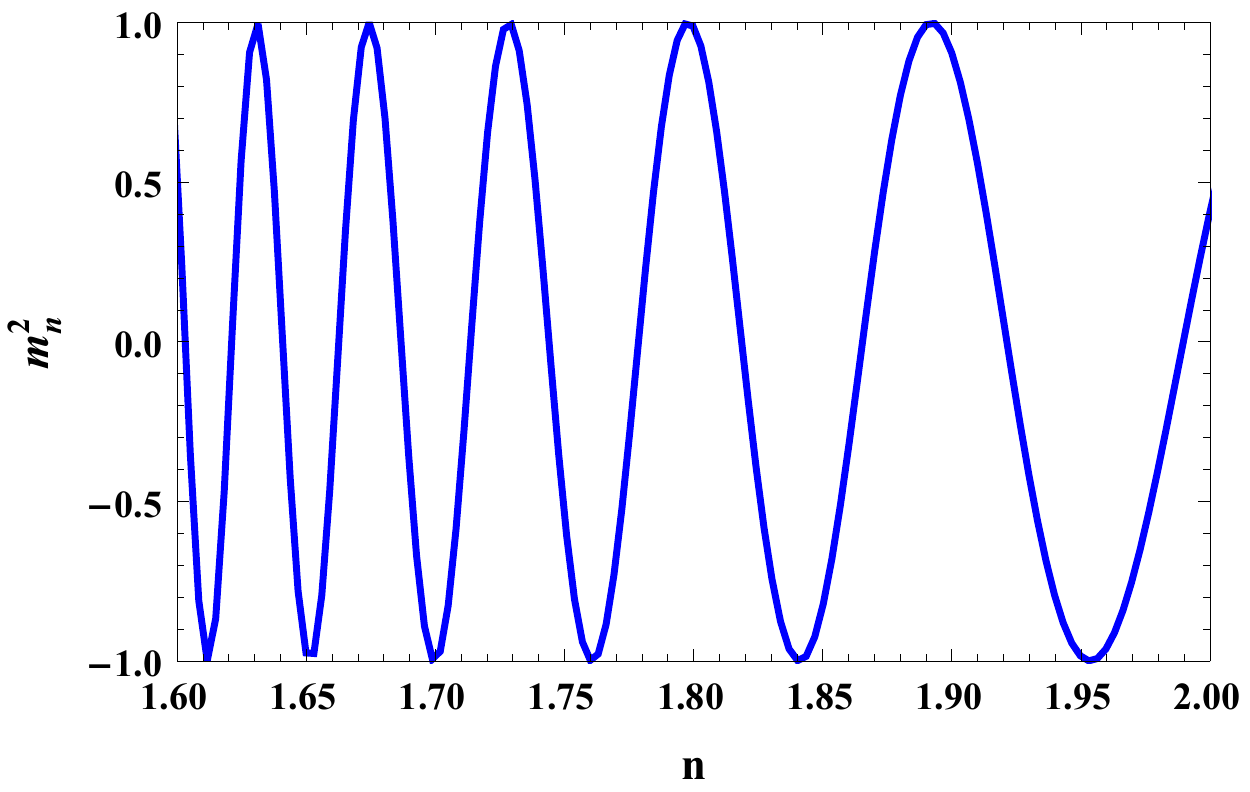}}
\caption{{\it{Top panel: the spectrum $m^{2}_{n}$ ($n$ is the number of modes) of the almost massless mode for the values $m^2=0.0419$  and $\gamma=0.1$. Bottom panel: the spectrum of the highly excited mode with $m^2=6$ and 
$\gamma=2$.}}}\label{UVRE6}
\label{planohwkhz}
\end{figure}

Fig. \ref{UVRE6} shows the spectrum of low and high altitude modes as we increase Horndeski-like's $\gamma$ parameter. For small values of $\gamma$, we keep all other modes of the analysis that interpolate between an AdS$_5$ and dS$_5$ pure. Thus, when we increase the value of $\gamma$, one mode decreases 
to mass $m^2$, while all other modes remain at $m^2=\mathcal{O}(\gamma)$. However, we have all the heavy modes of the AdS$_5$ and dS$_5$ analysis numerically and even a very light mode trapped. At the critical point, the trapped very light mode becomes a trapped massless graviton, on the other hand, the densely spaced excited modes become the continuum of KK modes \cite{Karch:2003em}.

\newpage
\section{Conclusions and discussion}\label{V4}

In this manuscript we considered Horndeski-like gravity with two scalar fields and we studied solutions which interpolate between asymptotically de Sitter and asymptotically Anti-de Sitter spacetimes. In particular, we developed the first-order formalism with two axionic fields, and we investigated
 three different cases, namely exponential, vacuum, and smooth superpotential solutions. With these solutions we have shown that a RG flow is established, and we obtained a turnaround in the warp factor in the braneworld scenario, which modifies gravity at long-distance scales. We mention that the model is free of ghosts and the matter sector that violates the $c$-theorem is physical, which is not the case in quasi-location models in which there is a ghost when four-dimensional gravity is reproduced and 
furthermore require non-physical matter to violate the $c$-theorem. 
 
Our construction has interesting implications for holography with regard to the AdS/CFT and dS/CFT scenarios \cite{Strominger:2001pn,Karch:2000gx,DeWolfe:1999cp}. In particular, the holographic description in our setup shows that a CFT resides on the disk which is the remainder of the true AdS$_{5}$ 
boundary. On the other hand, for dS$_{5}$, we have a boundary for the two-dimensional sphere of the minimum area that cuts the four-dimensional surface.
With the above, all brane physics has its information reduced to entanglement entropy at the common disk boundary and AdS$_{4}$-dS$_{4}$. However, such a description is not suitable for studying the local physics on the brane \cite{Karch:2000ct}, and thus one must divide the mass excitation into two sets, one double for a CFT in the true limit and the other for a CFT on the brane \cite{Alishahiha:2004md}. Hence, we produced a relation analogous to that arising in distorted compression or Randall-Sundrum geometries \cite{Randall:1999vf} with multiple throats \cite{Susskind:2021esx}. Furthermore, we have shown that near-zero-mode mass is attributed to the long-distance behavior of the warp factor that is sensitive to 
the $\gamma$ parameter of boundary physics.  

For the dS brane, the graviton is trapped as in the original (critical, Minkowski) Randall-Sundrum brane, and thus the volcano potential has a genuine zero mode, a feature that  is evident from  the
equations of   tensor perturbations. Additionally, for the AdS brane we found that the potential goes to infinity at the edges and, effectively, we acquire a box-type potential where the zero modes are 
eliminated and we need to deal with a massive graviton. Since the potential tends to a constant at infinity, we obtain  a massless graviton separated by a gap from the Kaluza-Klein (KK) tower. Finally, note that   in the dS case, we do not even need a brane, since dS traps a graviton ``in the central slice'' of the pode and antipode of our black hole ansatz with flat slicing (see \cite{Karch:2003em,Alishahiha:2004md} for further discussions for gravitons trapped in dS-type geometry).

\acknowledgments

We would like to thank Andreas  Karch, Edgar Shaghoulian and Mois\'es Bravo 
Gaete for fruitful discussions.  

\end{document}